\documentclass{article}

\usepackage{arxiv}
\usepackage[utf8]{inputenc} % allow utf-8 input
\usepackage[T1]{fontenc}    % use 8-bit T1 fonts
\usepackage{hyperref}       % hyperlinks
\usepackage{url}            % simple URL typesetting
\usepackage{booktabs}       % professional-quality tables
\usepackage{amsfonts}       % blackboard math symbols
\usepackage{amsmath}        % equation alignment
\usepackage{bm}
\usepackage{nicefrac}       % compact symbols for 1/2, etc.
\usepackage{microtype}      % microtypography
\usepackage{cleveref}       % smart cross-referencing
\usepackage{lipsum}         % Can be removed after putting your text content
\usepackage{graphicx}
\usepackage[square,sort,comma,numbers]{natbib}
\usepackage{doi}
\usepackage[labelfont=bf]{caption}
\usepackage{float,lscape}
\usepackage[nottoc,numbib]{tocbibind}
\usepackage{caption}        % To add source references
\usepackage[ruled,vlined]{algorithm2e}
\usepackage{subcaption}
\usepackage{scalerel}
\usepackage{xcolor}
\usepackage[width=.9\textwidth]{caption}

\title{Variational Bayes latent class analysis for EHR-based phenotyping with large real-world data}

\date{}

\author{
Brian Buckley
\thanks{School of Mathematics \& Statistics, University College Dublin, Ireland}\\
\texttt{brian.buckley.1@ucdconnect.ie}
\And
Adrian O'Hagan
\footnotemark[1] \thanks{The Insight Centre for Data Analytics, University College Dublin, Ireland}
\And
Marie Galligan
\thanks{School of Medicine, University College Dublin, Ireland}
}

\renewcommand{\headeright}{}
\renewcommand{\undertitle}{}
\hypersetup{
pdftitle={Variational Bayes latent class approach for EHR-based phenotyping},
pdfsubject={q-bio.NC, q-bio.QM},
pdfauthor={Brian Buckley, Adrian O'Hagan, Marie Galligan},
pdfkeywords={Big Bayes, Bayesian Latent Class Analysis, Variational Inference},
}

\begin{document}
\maketitle

\begin{abstract}
Bayesian approaches to patient phenotyping in clinical observational studies have been limited by the computational challenges associated with applying the Markov Chain Monte Carlo (MCMC) approach to real-world data.  Approximate Bayesian inference via optimization of the variational evidence lower bound, variational Bayes (VB), has been successfully demonstrated for other applications. We investigate the performance and characteristics of currently available VB and MCMC software to explore the practicability of available approaches and provide guidance for clinical practitioners. Two case studies are used to fully explore the methods covering a variety of real-world data. First, we use the publicly available Pima Indian diabetes data to comprehensively compare VB implementations of logistic regression. Second, a large real-world data set, Optum\textsuperscript{\texttrademark} EHR with approximately one million diabetes patients extended the analysis to large, highly unbalanced data containing discrete and continuous variables. A Bayesian patient phenotyping composite model incorporating latent class analysis (LCA) and regression was implemented with the second case study.  We find that several data characteristics common in clinical data, such as sparsity, significantly affect the posterior accuracy of automatic VB methods compared with conditionally conjugate mean-field methods. We find that for both models, automatic VB approaches require more effort and technical knowledge to set up for accurate posterior estimation and are very sensitive to stopping time compared with closed-form VB methods.  Our results indicate that the patient phenotyping composite Bayes model is more easily usable for real-world studies if Monte Carlo is replaced with VB. It can potentially become a uniquely useful tool for decision support, especially for rare diseases where gold-standard biomarker data is sparse but prior knowledge can be used to assist model diagnosis and may suggest when biomarker tests are warranted.
\end{abstract}

% keywords can be removed

\section{Introduction}

With the growing acceptance by clinical regulators of using real-world evidence to supplement clinical trials, there is increasing interest in the use of Bayesian analysis for both experimental and observational clinical studies~\citep{boulanger703and}.  The identification of specific patient groups sharing similar disease-related phenotypes is core to many clinical studies~\citep{he2023trends}.  Secondary use of electronic health records (EHR) data has long been incorporated in phenotypic studies~\citep{hripcsak2013next}.  Recent studies cover a wide range including health economics models~\citep{xia2022p30}, innovative trial designs for rare disease drug design~\citep{carlin2022bayesian}, and application of EHR information for antibiotic dose calculations~\citep{alsowaida2022vancomycin}.  The most popular approach that has evolved for identification of cohorts belonging to a particular phenotype of interest is an expert-driven rules-based heuristic dependent on data availability for a given phenotypic study e.g. Cuker et. al. (2023)~\citep{cuker2023early}.  The application of machine learning to patient phenotyping has become an active area of research~\citep{yang2023machine} but such approaches do not consider prior knowledge and hence are limited, particularly where data is sparse, for example with rare diseases. The patient phenotyping approach presented in this paper uses Bayesian statistics to provide a formal mathematical method for combining prior information with current information at a study design stage for identification of phenotypes as a means of selecting a study population.  This offers a mathematically principled method for incorporating expert opinion and other prior information that can potentially improve study population precision.

Markov Chain Monte Carlo (MCMC) is considered the gold standard for Bayesian inference because, in theory, it asymptotically converges to the true posterior distribution if run for a sufficient number of iterations thereby reaching a stationary posterior distribution ~\citep{geyer1992practical}.  Its application has been limited by the computational challenges of applying MCMC to large real-world clinical data. In Bayesian analysis, we are usually interested in the posterior distribution $p(Z|X)$ where $\mathbf{Z}$ are latent variables $\{z_1,...,z_n\}$ and $\mathbf{X}$ are observed data $\{x_1,...,x_n\}$.  In many real-world models of interest, the denominator of Bayes' theorem, $p(X)$, often referred to as the \textit{evidence}, and evaluates to $\int p(Z,X)dZ$, typically gives rise to analytically or computationally intractable integrals~\citep{murphy2012machine}.  Approximate Bayesian inference via optimization of the one-sided evidence lower bound is often called Variational Bayes (VB) or Variational Inference (VI).  In VB, optimization is used to find  an optimum distribution $q(Z)$ from a family of tractable distributions $\mathbf{Q}$ such that it is as close to the posterior distribution $p(Z|X)$ as possible. The optimization aims to locate $q^*(Z)$, the distribution that minimises the Kullback-Leibler (KL) divergence of $q(Z)$ and $p(Z|X)$.  This optimization approach is often significantly more computationally efficient than MCMC at the expense of an approximated posterior that cannot be improved with more iterations.

VB has been successfully demonstrated for other patient phenotyping applications.  Song et al. applied variational inference with a deep learning natural language processing (NLP) approach to patient phenotyping~\citep{song2022automatic}.  Li et al. propose another NLP model, \textit{MixEHR}, that applies a latent topic model to EHR data~\citep{li2020inferring}.  They apply a VB coordinate ascent approach to impute mixed disease memberships and latent disease topics that apply to a given patient EHR cohort.  Hughes et al. developed a mean-field VB model for multivariate generalized linear mixed models applied to longitudinal clinical studies~\citep{hughes2023fast}.

Latent Class Analysis (LCA) is widely used when we want to identify patient phenotypes or subgroups given multivariate data~\citep{lanza2013latent}.  A challenge in biomedical LCA is the prevalence of mixed data, where we may have combinations of continuous, nominal, ordinal and count data, further complicated by missing and inaccurate data across variables.  Bayesian approaches to LCA may better account for this data complexity. The Bayesian approach serves as a connection between rule-based phenotypes, which heavily depend on medical expert knowledge and opinion, and data-driven machine learning methods that solely rely on the information within the data being analyzed, without the ability to incorporate relevant prior knowledge ~\citep{mcparland2016model}.  White and Murphy (2014) included a variational inference approach to latent class analysis with their \textit{BayesLCA} R package~\citep{white2014bayeslca}.

Hubbard et al.~\citep{hubbard2019bayesian} proposed a composite LCA/regression model that might be used in a general context for observational studies that use EHR data. They consider the common clinical context where gold-standard phenotype information, such as genetic and laboratory data, is not fully available. A general model of this form has high potential applicability for use in clinical decision support across disease areas and with primary and secondary clinical databases. In this paper we want to evaluate VB for this context and compare with several MCMC implementations to determine which are potentially suitable choices for large real-world biomedical studies of this form.  Our primary objective is to determine if VB can enable wider use of Bayesian approaches in large biomedical studies with acceptable accuracy in a clinical setting.

This study is motivated by a patient phenotyping composite Bayes LCA/regression model of the form proposed by Hubbard et al.~\citep{hubbard2019bayesian}. It can potentially become a uniquely useful tool for clinical decision support, especially for rare disease areas where gold-standard biomarker data is sparse, if the computational challenges of MCMC can be ameliorated.  In this paper, we use the same composite Bayesian LCA/regression model form and motivating example as Hubbard et al., namely pediatric type 2 diabetes mellitus (T2DM), in order to test if the proposed LCA model translates to a different EHR database with the same target disease under study (Optum\textsuperscript{\texttrademark} EHR).  Pediatric T2DM is rare so it naturally gives rise to the data quality issues we would like to explore with our approach.  We extend the Hubbard et al. work to incorporate VB to investigate if a much larger EHR data set is amenable to this form of Bayesian model.  Our aim in this work is to investigate if a VB approach to Bayesian LCA/regression phenotyping scales to real-world large EHR data and delivers a posterior approximation acceptably close to MCMC in a clinical setting.  We compare a range of readily available VB software within two case studies comparing how closed-form and black-box VB methods perform with clinical data.  The structure of the paper is as follows. Section ~\ref{sec1} details the VB algorithms and available software compared including comparison with a maximum likelihood estimation (MLE) approach for completeness since the majority of clinical analyses currently employ MLE. Note that the aim of this paper is to focus on scalable Bayesian methods so we did not carry out a detailed comparison with either machine learning or frequentist/MLE approaches. This section also describes the data used for each case study and how each serves our primary objective of comparing a wide range of VB algorithms and software with a special focus on generalizable Bayesian models for large real-world data; Section ~\ref{sec2} presents results for both use cases; Section ~\ref{sec3} concludes with a summary and discussion of advantages and challenges of using VB for EHR-based patient phenotyping with suggestions for further work.  The primary goal is matching the gold-standard MCMC algorithm as closely as possible while enabling timely completion of large-scale clinical observational studies.

\section{Materials and Methods}\label{sec1}

Our primary objective is to investigate if a VB approach to a potentially generalizable composite Bayesian LCA/regression phenotyping model scales to real-world large EHR data and delivers a posterior approximation acceptably close to MCMC in a clinical setting.  We find there is a lack of available VB software suitable for our composite Bayesian LCA/regression model. For that reason, we included a simpler case study using a purely logistic regression model with the widely available Pima Indians' diabetes data \footnote{https://www.kaggle.com/datasets/uciml/pima-indians-diabetes-database}.  This enabled us to include conditionally conjugate closed-form methods that are available for VB logistic regression in a wider comparison of VB and MCMC.

MCMC has become the \textit{de facto} gold-standard method for Bayesian analysis due its theoretical guarantee that, if iterated to infinity, it will eventually sample from the true posterior distribution given by Bayes' rule\mbox{~\citep{metropolis1949monte}}. In practice, it is difficult to determine how many iterations are needed before an MCMC sampling chain has converged to stable bounds around the true posterior.  We therefore usually run several separate sampling chains for some pre-specified number of iterations, including so-called burn-in initial iterations that we ignore for convergence assessment\mbox{~\citep{resnik2010gibbs}}.  Convergence assessment usually includes comparison of diagnostic measures between the separate sampling chains\mbox{~\citep{roy2020convergence}}. \textit{JAGS} is a Metropolis-Hastings sampler that specializes in Gibbs sampling\mbox{~\citep{plummer2003jags}}.  Gibbs sampling can sample discrete as well as continuous latent parameters.  \textit{Stan} is a Metropolis-Hastings sampler that uses a different approach, Hamiltonian Monte Carlo, that can be faster to reach convergence but, instead of sampling discrete latent variables, integrates across each class of each discrete latent variable\mbox{~\citep{carpenter2017stan}}.  This is a technique called marginalizing-out or integrating-out the discrete latent variables from the posterior distribution \footnote{See section 8: Latent Discrete Parameters in the Stan Users Guide}. A benefit of marginalizing-out discrete latent variables is a model specification that allows greater exploration of the tails of the distribution (see footnote 2).  However, it requires specific \textit{Stan} model syntax which complicates translation of JAGS models with discrete parameters.

VB has become popular in cases where MCMC computation is particularly challenging. VB employs optimization of the lower bound on the marginal likelihood to minimise the Kullback-Leibler divergence of a member of some posited family of approximations to the real posterior distribution~\citep{jordan1999introduction}.  This lower bound is often referred to as the Evidence Lower Bound (ELBO) or the Bayes free energy. The Blei et. al. (2018) comprehensive review of VB~\citep{blei2017variational}, addresses VB from a statistical viewpoint. Their paper covers both the deterministic mean-field approach, coordinate ascent variational inference (CAVI) as well as stochastic mean-field optimization that scales to very large data. They provide outline algorithms for both approaches. Stochastic Variational Inference (SVI) makes use of repeated subsampling of the data points to noisily optimize the ELBO~\citep{hoffman2013stochastic}.

For the logistic regression case study, we compared \textit{JAGS} and \textit{Stan} MCMC with ten VB software packages plus a textbook implementation (Table ~\ref{tab:vb_algos}).  Five of the software packages use automatic black-box approaches.  Automatic approaches focus on optimizing difficult variational objectives where the exponential family property for the proposed variational density may not apply~\citep{kucukelbir2017automatic}.  These methods attempt to generalize nonconjugate inference in order to apply across many models without specific model derivation of the objective functions, hence the "black-box" label. For example, Automatic Differentiation Variational Inference (ADVI)~\citep{kucukelbir2017automatic} is a hybrid approach using Monte Carlo estimates of the ELBO gradient written as an expectation that can be repeatedly sampled within a stochastic optimization algorithm. ADVI transforms the support of latent variables to the real coordinate space before computing the ELBO using Monte Carlo integration.  It then uses stochastic gradient ascent to maximize the ELBO.  All of this is performed automatically, without user input.

\clearpage
\begin{center}
\captionof{table}{VB and MCMC algorithms in the logistic regression case study.  Automatic VB methods do not require analytical derivation of the ELBO objective function.}
\vspace{3mm}
\scalebox{0.80}{
\begin{tabular}{l l l l l}
\toprule
\textbf{Algorithm} & \textbf{Description} & \textbf{Type} & \textbf{Automatic} & \textbf{Programming} \\
\midrule
CAVI & Coordinate Ascent Variational Inference  & mean-field VB & No & R \\
Own VI & Textbook implementation of CAVI based on~\citep{murphy2012machine} & mean-field VB & No & R \\
SVI & Stochastic Variational Inference  & mean-field VB & No & R \\
varbvs & Fast Variable Selection for Large-scale Regression & mean-field VB & No & R \\
sparsevb & Spike-and-Slab VB for Linear and Logistic Regression & mean-field VB & No & R \\
Stan MC & Stan Hamiltonian Monte Carlo & MCMC & No & R (rstan) \\
Stan VB & Stan ADVI Variational Bayes & mean-field or full-rank VB & Yes & R (rstan, CmdStanR) \\
ADVI & Automatic Differentiation Variational Inference & mean-field VB & Yes & Python \\
FRADVI & Full-rank Automatic Differentiation Variational Inference  & full-rank VB & Yes & Python \\
NFVI &  Normalizing Flow Variational Inference & mean-field or full-rank VB & Yes & Python \\
ASVGD & Amortized Stein Variational Gradient Descent & operator VB & Yes & Python \\
\bottomrule
\label{tab:vb_algos}
\end{tabular}}
\end{center}

The \textit{Stan} implementation for ADVI incorporates several algorithm hyperparameters that can be tuned \footnote{See section 13: Variational Inference Algorithm: ADVI in the Stan User's Guide}.  We found combinations of several algorithm hyperparameters affected both runtime and how closely the result converged (or failed to converge) to estimates similar to those obtained with MCMC.  The \textit{iter} hyperparameter is important to ensure the algorithm is allowed to run long enough to achieve ELBO convergence.  For our EHR data set, we found a large number of iterations were needed (15,000). The default of 10,000 seems a good general starting point. We found the \textit{elbo\_samples} default (100) was far too low for our EHR data. The \textit{elbo\_samples} hyperparameter sets the number of samples for Monte Carlo estimate of the ELBO.  For our data we found best accuracy with a value around 10,000.  This setting increased the run time of the algorithm to the extent it was not significantly quicker than MCMC. The \textit{tol\_rel\_obj} hyperparameter sets the stopping criterion.  For our EHR data, we set it lower (0.001) than the default of 0.01 as we found cases where the ELBO had not converged at the time the algorithm stopped.  We used the default values for other hyperparameters.

For the Bayesian composite LCA/regression case study, we first reproduced the Hubbard et al. model~\citep{hubbard2019bayesian} with \textit{JAGS} MCMC~\citep{ plummer2003jags} using the Optum\textsuperscript{\texttrademark} EHR T2DM data.  This gives us confidence that the composite Bayes LCA/regression model can generalize to other similar EHR data.  We then followed up with \textit{Stan}~\citep{carpenter2017stan} implementations of Hamiltonian MCMC and VB to investigate how the model specification translates to other algorithmic approaches.  As we could not find available closed-form implementations suitable for our composite Bayes LCA/regression model, such as coordinate ascent or stochastic mean-field VB, we used \textit{Stan} Automatic Differentiation Variational Inference (ADVI)~\citep{kucukelbir2015automatic} for this case study.

\subsection{Logistic Regression Model}

We used the standard logistic regression model for binary classification.  Given a data set with $n$ observations and $m$ observed feature variables, where $x_i$ represents the observed variables for the $i^th$ observation, and $y_i$ represents the binary response, the logistic regression model predicts the probability $p(y_i = 1|x_i)$ using the logistic function

\begin{align}
p(y_i = 1|x_i) = \frac{1}{1 + e^{-z_i}}
\end{align}

where $z_i$ is the linear combination of the features and their corresponding coefficients:

\begin{align}
z_i = \beta_0 + \beta_1x_{i1} + \beta_2x_{i2} + \dots + \beta_mx_{im}
\end{align}

where $\beta_0$ is the intercept coefficient and we have $\beta_m$ coefficients for the $1, \dots, m$ feature variables.  In Bayesian logic regression we can define priors on the coefficients.  We used normal priors in this study as all predictor variables are continuous.

\subsubsection{Pima Indians' Diabetes Data}

We used the Kaggle subset\footnote{https://www.kaggle.com/datasets/uciml/pima-indians-diabetes-database} of Pima Indian adult T2DM data to facilitate a wide comparison of VB implementations~\citep{smith1988using}.  The response, $y_i$, is the variable \textit{Outcome} (diagnosed T2DM).  The predictors, $x_i$, are all continuous variables (Pregnancies, Glucose, Blood Pressure, Skin Thickness, Insulin, BMI, Diabetes Pedigree and Age).  A logistic regression model was fitted to all predictors as there are many available VB software implementations for logistic regression so it serves as a suitable comparative baseline approach.  Our objective with this case study is to compare conditionally conjugate closed-form methods e.g.~\citep{durante2019conditionally} with black-box automatic methods e.g.~\citep{kucukelbir2017automatic} using clinically relevant data.

\subsubsection{VB Logistic Regression Software}

Two VB software libraries in R were compared; \textit{sparsevb}~\citep{ray2022variational} and \textit{varbvs}~\citep{carbonetto2017varbvs} along with published implementations of CAVI and SVI from the GitHub \footnote{https://github.com/tommasorigon/logisticVB} of Durante \& Rigon (2019)~\citep{durante2019conditionally}. \textit{sparsevb} uses mean-field CAVI VB.  Their package focuses specifically on variable selection in high dimensional data for regression models using a spike-and-slab prior to induce sparsity.  Their package includes a novel parameter updating order to improve the CAVI algorithm performance.  \textit{varbvs} is also focused on variable selection, in this case aimed at genome-wide association studies.  \textit{varbvs} is sensitive to the variable ordering and initialization of optimization procedure. The Python library \textit{PyMC3}~\citep{salvatier2016probabilistic} supports four VB methods that were included in this study (Table ~\ref{tab:vb_algos}). All four are automatic black-box methods similar to those found in \textit{Stan}. The ADVI algorithm from \textit{PyMC3} has a similar programming approach to \textit{Stan}. Full-rank ADVI (FRADVI) generalizes the mean-field approximation. The off-diagonal terms in the covariance matrix capture posterior correlations across latent random variables.  This should result in a more accurate posterior approximation, though we found otherwise for Pima Indians' diabetes data (Figure 4).  The Normalizing Flows VI (NFVI) approach originates from Rezende and Mohamed (2015)~\citep{rezende2015variational} whereby a standard initial density is transformed into a more complex proposed density by applying a sequence of invertible transformations until a preset level of complexity is obtained. The aim is to improve the posterior accuracy. The Amortized Stein Variational Gradient Descent (ASVGD) approach is from Liu and Stein (2016)~\citep{liu2016stein}.  ASVGD is related to the work by Rezende and Mohamed (2015). They introduce a particle-based framework that allows for automatic differentiation through stochastic computation graphs to efficiently estimate gradients. In this approach, the mean parameters are treated as a set of particles. We included a na\"{\i}ve textbook implementation of CAVI from Murphy (2012)~\citep{murphy2012machine} to assess how better the Durante \& Rigon published implementation is.  We applied 5-fold cross validation (CV) to the data set to investigate the stability of the different implementations.  For algorithms containing hyperparameters, we performed a grid search over a range for each hyperparameter within each CV iteration.

\subsection{Composite Bayes LCA/Regression Model}

The canonical form for LCA assumes that each observation $i$, with $\textbf{U}$ observed categorical response variables $\textbf{x}_i = \{x_{i1}, x_{i2}, …, x_{iU}\}$,  belongs to one of C latent classes. Class membership is represented by a latent class indicator $z_i = \{1,...,C\}$.  The marginal response probabilities are

\begin{align}
P(\textbf{x}_i) = \sum\limits_{c=1}^C\pi_c\prod\limits_{u=1}^UP(x_{iu}|z_i=c)
\end{align}

Where $z_i$ is the latent class that observation $i$ belongs to and $\boldsymbol{\pi}_c$ is the probability of being in class $c$.  The variables $\textbf{x}$ are assumed to be conditionally independent given class membership, an assumption known as local independence.

The composite Bayes LCA/regression model applied is from Hubbard et al.~\citep{hubbard2019bayesian} and follows the general specification shown in Table~\ref{table:modelSpec}.  In the T2DM model example, $D_i$ represents the latent dichotomous propensity for observation $i$ to be in the T2DM class i.e. the value of $\textbf{C}$ in eq(3) denotes that either the patient has T2DM or does not have T2DM. There are three variables in $\textbf{U}$; biomarker indicator $R_{ij}$, where j has two categories, laboratory test availability for HbA1c and Random Blood Glucose for observation $i$; ICD clinical codes $W_{ik}$, where k has two categories, diagnosis of T2DM and at least one endocrinologist visit; and diabetes medications $P_{il}$, where l has two categories, metformin and insulin.  The latent phenotype variable for each patient, $D_i$, is assumed to be associated with patient characteristics that include demographics (for pediatric T2DM they are age, higher-risk ethnicity and BMI z-score) denoted by $\textbf{X}$ in the model specification.  This model allows for any number of clinical codes or medications and in this model, each clinical code and medication are binary indicator variables to specify if the code or medication was present for that patient.  Since it is common for biomarker laboratory tests, $Y_{ij}$, to be missing across a cohort of patients, biomarker availability, $R_{ij}$, is an important component of the model.  This is because biomarkers are widely considered to be high quality prognostic data for many disease areas~\citep{hadjadj2004prognostic} and availability of a biomarker measurement can be predictive of the phenotype since laboratory tests are often proxies for physician diagnoses.

\clearpage
\begin{center}
\captionof{table}{Model specification for Bayesian latent variable model for EHR-derived phenotypes for patient $i$.\label{table:modelSpec}}
\vspace{5mm}
\scalebox{0.80}{
\begin{tabular}{lllll}
\toprule
                                        & \multicolumn{1}{c}{\bfseries Variable} & \multicolumn{1}{c}{\bfseries Model} & \multicolumn{1}{c}{\bfseries Priors}   \\
\midrule
                                        &                       &                                                                &            \\
\textbf{Latent Phenotype}               & $D_i$                 & $D_i \sim $Bern$(g(\bm{X}_i\bm{\beta}^D + \eta_i))$            & $\beta^D \sim $MVN$(0, \Sigma_D); \eta_i \sim $Unif$(a,b)$              \\
                                        &                       &                                                                &            \\
\textbf{Availability of Biomarkers}     & $R_{ij}, j=1,...,J$   & $R_{ij} \sim $Bern$(g((1,\bm{X}_i,D_i)\bm{\beta}^R_j))$        & $\beta^R_j \sim $MVN$(\mu_R, \Sigma_R)$                                  \\
                                        &                       &                                                                &            \\
\textbf{Biomarkers}                     & $Y_{ij}, j=1,...,J$   & $Y_{ij} \sim $N$(g((1,\bm{X}_i,D_i)\bm{\beta}^Y_j, \tau^2_j))$ & $\beta^Y_j \sim $MVN$(\mu_Y, \Sigma_Y); \tau^2_j \sim $InvGamma$(c,d)$  \\
                                        &                       &                                                                &            \\
\textbf{Clinical Codes}                 & $W_{ik}, k=1,...,K$   & $W_{ik} \sim $Bern$(g((1,\bm{X}_i,D_i)\bm{\beta}^W_k))$        & $\beta^W_k \sim $MVN$(\mu_W, \Sigma_W)$                                  \\
                                        &                       &                                                                &            \\
\textbf{Prescription Medications}       & $P_{il}, l=1,...,L$   & $P_{il} \sim $Bern$(g((1,\bm{X}_i,D_i)\bm{\beta}^P_l))$        & $\beta^P_l \sim $MVN$(\mu_P, \Sigma_P)$                                  \\
                                        &                       &                                                                &            \\
\bottomrule
                                        &                       &                                                                &            \\
                                        &                       &       & \multicolumn{1}{c}{$\bm{ g(\bm{\cdot}) = exp(\bm{\cdot})/(1 + exp(\bm{\cdot})) }$} \\
\end{tabular}}
\label{table:modelSpec}
\end{center}

The model likelihood for the $i^{th}$ patient is given by

\begin{align}
\mathcal{L}(\eta_i,\beta^D,\beta^R,\beta^Y,\beta^W,\beta^P,\tau^2|X_i) & = \sum_{d=0,1}{P(D_i=d|\eta_i,\beta^D,X_i)} \nonumber \\
& \prod_{j=1}^{J}{f(R_{ij}|D_i=d,X_i,\beta_j^R)f(Y_{ij}|D_i=d,X_i,\beta_j^Y,\tau_j^2)^{R_{ij}}} \nonumber \\
& \prod_{k=1}^{K}{f(W_{ik}|D_i=d,X_i,\beta_k^W)} \prod_{l=1}^{L}{f(P_{il}|D_i=d,X_i,\beta_i^P)} \label{eqn:likelihood} \\ \nonumber
\end{align}

where $\beta^D$ associates the latent phenotype to patient characteristics, $\eta_i$ is a patient-specific random effect parameter and parameters $\beta^R,\beta^Y,\beta^W$ and $\beta^P$ associate the latent phenotype and patient characteristics to biomarker availability, biomarker values, clinical codes, and medications respectively.  $X_i$ represents $M$ patient covariates, such as demographics ($X_i = X_{i1},...,X_{iM}$).  The mean biomarker values are shifted by a regression quantity $\beta^Y_{j,M+1}$ for patients with the phenotype compared to those without \footnote{$M+1$ is to account for the regression intercept}. The sensitivity and specificity of binary indicators for clinical codes, medications and the presence of biomarkers are given by combinations of regression parameters. For instance, in a model with no patient covariates, sensitivity of the $k^{th}$ clinical code is given by $expit(\beta^W_{k_0} + \beta^W_{k_1})$, while specificity is given by $1-expit(\beta^W_{k_0})$, where $expit(\cdot) = exp(\cdot)/(1+exp(\cdot))$.  We validate this model with a real-world example, namely pediatric T2DM.  Table~\ref{table:modelExample} indicates how the composite Bayes LCA/regression phenotyping model maps to this disease area. $f(\bm{\cdot})$ in eq(~\ref{eqn:likelihood}) represents the probability function that will depend on the specific disease application of the model.

\clearpage
\begin{center}
\captionof{table}{Mapping the composite Bayes LCA/regression phenotyping model factors to the pediatric T2DM study for patient $i$. }
\vspace{3mm}
\scalebox{0.80}{
\begin{tabular}{lllll}
\toprule
                                        & \multicolumn{1}{c}{\bfseries Model Variable} & \multicolumn{1}{c}{\bfseries Data Elements}                \\
\midrule
                                        &                       &                                                                                   \\
\textbf{Latent Phenotype}               & $D_i$                 & Presence of T2DM for observation $i$ (binary latent variable, $D_i \in \{0,1\}$,  \\
                                        &                       & where $1$ indicates presence of T2DM) based on patient characteristics            \\
                                        &                       & e.g. demographics                                                                 \\
                                        &                       &                                                                                   \\
\textbf{Availability of Biomarkers}     & $R_{ij}, j=1,...,J$   & Availability of ($j$=1) Glucose and ($j$=2) HbA1c biomarker data                  \\
                                        &                       &                                                                                   \\
\textbf{Biomarkers}                     & $Y_{ij}, j=1,...,J$   & Laboratory test values for ($j$=1) Glucose and ($j$=2) HbA1c  CPT codes           \\
                                        &                       &                                                                                   \\
\textbf{Clinical Codes}                 & $W_{ik}, k=1,...,K$   & ($k$=1) ICD Code for T2 Diabetes and ($k$=2) CPT code for Endocrinologist Visit   \\
                                        &                       &                                                                                   \\
\textbf{Prescription Medications}       & $P_{il}, l=1,...,L$   & NCD Medication codes for ($l$=1) Insulin and ($l$=2) Metformin                    \\

                                        &                       &                                                                                   \\
\bottomrule
\label{table:modelExample}
\end{tabular}}
\end{center}

Informative priors are used to encode known information on the predictive accuracy of Glucose; $Y_{i1} = \mathcal{N}(90.6 + 42D_i, 16.93)$ and HbA1c; $Y_{i2} = \mathcal{N}(5.4 + 1D_i, 0.45)$ laboratory tests for T2DM.  The biomarker priors inform a receiver operating characteristic (ROC) model since the biomarkers are normally distributed. The prior values were selected to correspond to an area under the ROC curve (AUC) of 0.95~\citep{hubbard2019bayesian}.

Although the pediatric T2DM example comprises two measured variables for each of $Y_i$, $W_i$ and $R_i$, the model can be expanded to multiple biomarkers, clinical codes and medications for other disease areas and easily extends to include additional elements.  For example, the number of hospital visits or the number of medication prescriptions, or, for other disease conditions or study outcomes, comorbidities and medical costs, etc. Given our objective is to determine the utility and performance of VB in tackling a problem of this nature, in comparison to MCMC, we extend the Hubbard et al. work by comparing a range of alternative methods to \textit{JAGS} that included alternative MCMC and VB using \textit{Stan}. Our aims were to characterise the advantages and challenges posed by this extendable patient phenotyping composite Bayes LCA/regression model using VB approaches compared to the traditional Gibbs/Metropolis-Hastings sampling MCMC approach.  The following sub-sections discuss each of these approaches in turn.

\subsubsection{Optum\textsuperscript{\texttrademark} EHR Data}

We used licensed Optum\textsuperscript{\texttrademark} EHR data for the generalizable Bayesian LCA model.  This data set is comprised of EHR records from hospitals, hospital networks, general practice offices and specialist clinical providers across the United States of America.  It includes anonymized patient demographics, hospitalizations, laboratory tests and results, in-patient and prescribed medications, procedures, observations and diagnoses.  The data provides most information collected during a patient journey provided all care sites for a given patient are included in the list of Optum\textsuperscript{\texttrademark} data contributors.  This data set is one of the most comprehensive EHR databases in the world~\citep{wallace2014optum} and is used extensively for real-world clinical studies~\citep{dagenais2022use}.  The data set used in our case study contains Optum\textsuperscript{\texttrademark} EHR records collected between January 2010 and January 2020.

\begin{figure}[hbt!]
\centerline{\includegraphics[width=0.8\linewidth]{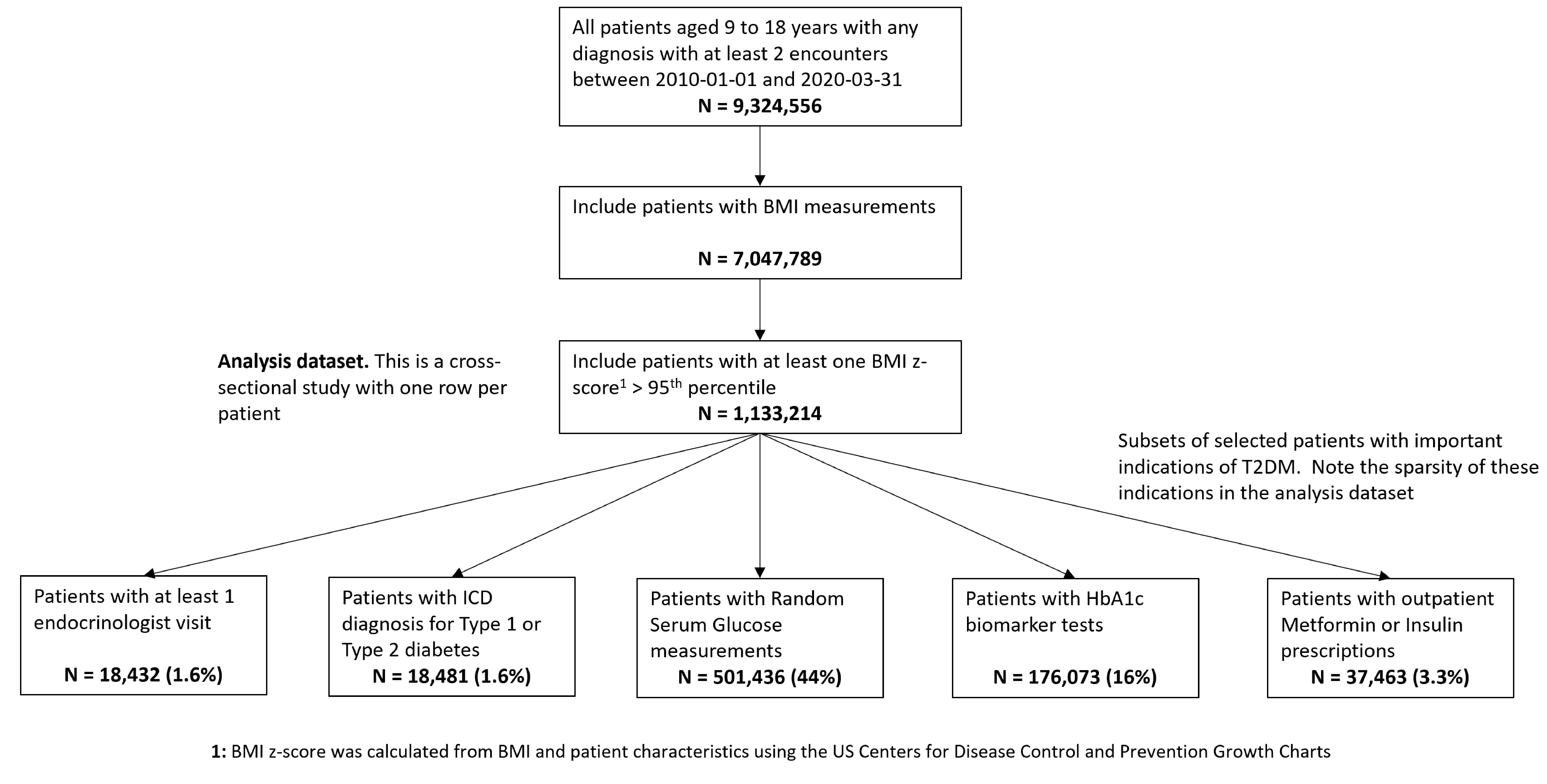}}
\caption{Data specification for pediatric patients at risk of T2DM in Optum\textsuperscript{\texttrademark} EHR database.  The bottom row shows the number of patients having the model study characteristics for the pediatric T2DM phenotype for the relevant variables.
\label{fig:attrition}}
\end{figure}

The initial processing step is to identify, from the overall Optum\textsuperscript{\texttrademark} data set, a cohort of pediatric patients with elevated risk of T2DM, in order to subsequently perform phenotyping.  For this analysis we extracted a patient cohort of pediatric patients with equivalent T2DM risk characteristics defined in Hubbard et al.~\citep{hubbard2019bayesian}.  We transformed our Optum\textsuperscript{\texttrademark} data schema into the same form as the Hubbard et al. data schema in order to test how well the proposed Bayesian LCA model translates from the Hubbard et al. PEDSnet EHR data to the Optum\textsuperscript{\texttrademark} EHR data. The PEDSnet data used by Hubbard et al. are located within the USA Northeast region which comprises about 13\% of the Optum\textsuperscript{\texttrademark} data which covers the whole USA.  To account for potential variance of pediatric T2DM prevalence across the USA we used the Northeast subset of the Optum\textsuperscript{\texttrademark} EHR for the composite Bayes LCA/regression comparison, although we also ran the study using all of the data.  The data specification in Figure~\ref{fig:attrition} shows how the data was extracted from Optum\textsuperscript{\texttrademark} EHR.  The overall objective for this specification was to arrive at the same patient identification rules used by Hubbard et al.  We also restricted the data variables to those used in the Hubbard et al. study.  The patient BMI z-score was calculated from the US Centers for Disease Control and Prevention Growth Charts~\citep{kuczmarski20022000} using patient age, sex and BMI. The full US Optum\textsuperscript{\texttrademark} EHR pediatric T2DM data extraction is compared with Hubbard et al. PEDSnet data in the supporting information (Table S1).

%\begin{figure}[hbt!]
%\centerline{\includegraphics[width=0.8\linewidth]{figs/Figure1.pdf}}
%\caption{Data specification for pediatric patients at risk of T2DM in Optum\textsuperscript{\texttrademark} EHR database.  The bottom row shows the number of patients having the model study characteristics for the pediatric T2DM phenotype for the relevant variables.\label{fig:attrition}}
%\end{figure}

\subsubsection{Gibbs Monte Carlo Sampling}

We used \textit{JAGS}~\citep{plummer2003jags} as the baseline approach for comparison with all other methods since it uses the simplest method of Metropolis-based MCMC~\citep{geman1984stochastic}.  Our initial baseline objective was to reproduce the Hubbard et al. model using the same methods they employed but against a different EHR database.  We used the same JAGS LCA model published by Hubbard et al. in their GitHub supplement \footnote{https://github.com/rhubb/Latent-phenotype/}.

\subsubsection{Hamiltonian Monte Carlo Sampling}

We used \textit{Stan} MC~\citep{carpenter2017stan} for an alternative MCMC comparison given that \textit{Stan} makes use of a different Metropolis-based sampling method to \textit{JAGS} and particularly for our model  employs marginalization of discrete latent variables. \textit{Stan} uses a variant of Hamiltonian Monte Carlo called 'No U-turn Sampler (NUTS)'~\citep{hoffman2014no}.  This approach includes a gradient optimization step so it cannot sample discrete latent parameters in the way JAGS can.  Instead, \textit{Stan} MC integrates the posterior distribution over the discrete classes~\citep{yackulic2020need}, so this is a useful comparison to the discrete sampling Gibbs approach.  We translated the \textit{JAGS} model directly to a \textit{Stan} model using similar sampling notation and had reasonable results (Table ~\ref{table:allResults}) though \textit{Stan} does have various helper functions e.g. log-sum-exp and the target log-probability increment statement.

\subsubsection{Variational Inference}

We used \textit{Stan} Automatic Differentiation VB~\citep{kucukelbir2017automatic} for comparison with MCMC.  In \textit{Stan}, there is only one implementation of variational inference, the automatic differentiation approach~\citep{kucukelbir2015automatic}.  We found this was challenging for the composite Bayes LCA/regression model.  For both \textit{Stan} MCMC and VB, we used the same \textit{Stan} model definition \footnote{See supplementary material Appendix S2 for the \textit{Stan} LCA model details}.  There is a lack of closed-form solutions implemented in R or Python for variational Bayes LCA.

\begin{figure}[hbt!]
\centering
  \includegraphics[width=0.6\linewidth]{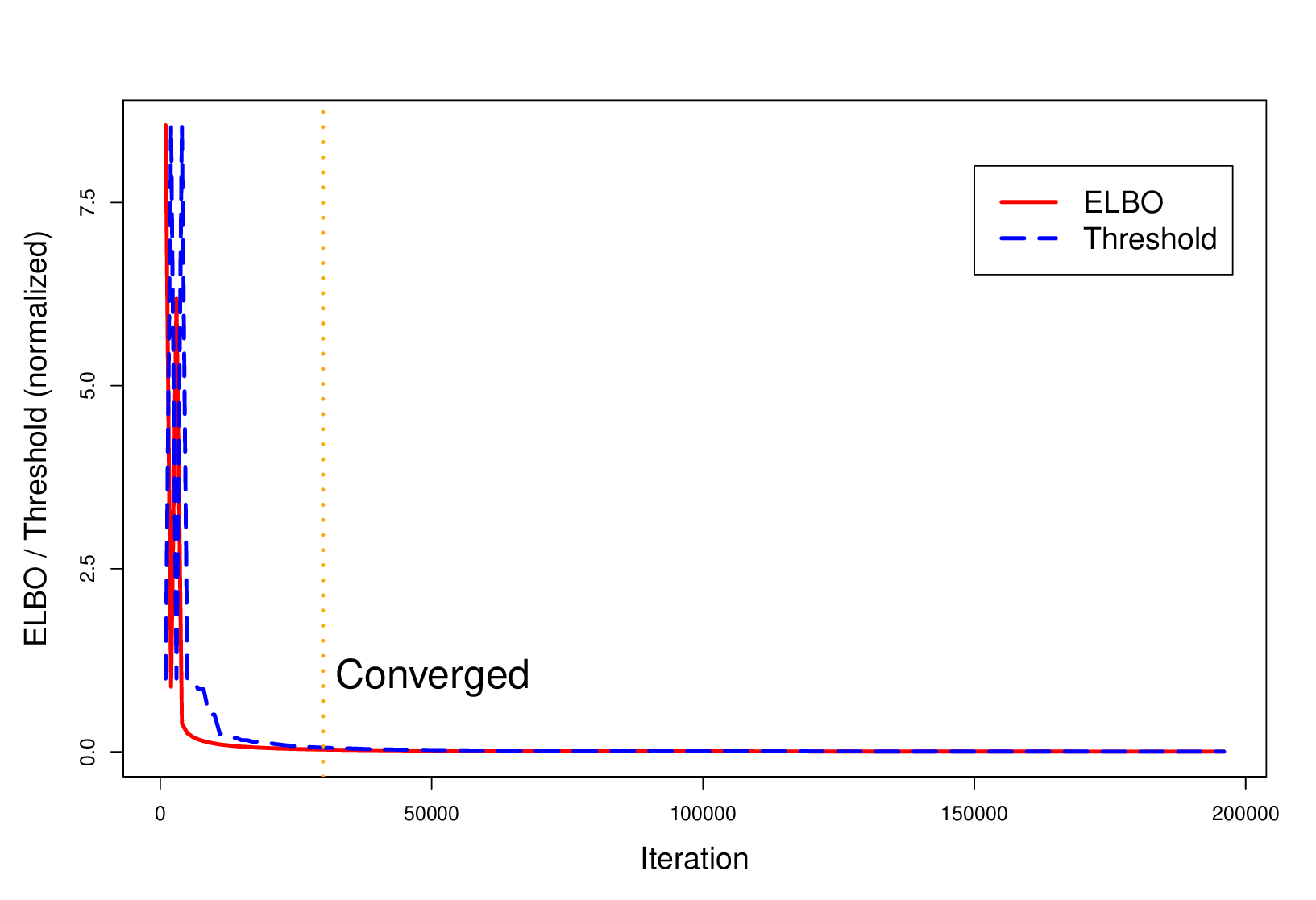}
\caption{Runtime ELBO (solid) and threshold delta (dashed) for all iterations.  ELBO and threshold have been normalized to the same scale.  We can see there is a diminishing return after about 30,000 iterations (vertical dotted line).  In this example, the algorithm ran for over a day longer than it needed to (on a Dell XPS computer with 8-cpu Intel core i9 and 64GB RAM memory) in finding the best posterior estimate it could generate.}
\label{fig:elboplot}
\end{figure}

Figure~\ref{fig:elboplot} illustrates a common problem with current VB implementations.  There are two hyperparameters for stopping, a maximum number of ELBO calculation iterations and a difference threshold (delta) between iteration $t$ and $t-1$.  In Figure~\ref{fig:elboplot} we can see that the $argmin \{ELBO\}$ has effectively converged after approximately 30,000 iterations but we have set maximum iterations to 200,000 and the threshold tolerance relatively low meaning it is never reached so the algorithm continues well past the best estimate it can produce.  Unfortunately, there is no way \textit{a priori} to know what a suitable threshold tolerance should be as ELBO values are unbounded or the effective number of iterations for ELBO convergence as this depends on various factors including model specification and algorithm tuning.  However, the variational approach uses significantly less computer memory than MCMC.  With this EHR phenotyping model on our computer system VB used 1.8GB versus 37GB of RAM memory for a three-chain MCMC.

%\begin{figure}[hbt!]
%\centering
%  \includegraphics[width=0.6\linewidth]{figs/Figure2.pdf}
%\caption{Runtime ELBO (solid) and threshold delta (dashed) for all iterations.  ELBO and threshold have been normalized to the same scale.  We can see there is a diminishing return after about 30,000 iterations (vertical dotted line).  In this example, the algorithm ran for over a day longer than it needed to (on a Dell XPS computer with 8-cpu Intel core i9 and 64GB RAM memory) in finding the best posterior estimate it could generate.}
%\label{fig:elboplot}
%\end{figure}

\subsubsection{Comparison with Maximum Likelihood Approach}

We used the R package \textit{clustMD}~\citep{mcparland2016model} to compare a maximum likelihood (MLE) clustering approach to Bayesian LCA.  \textit{clustMD} employs a mixture of Gaussian distributions to model the latent variable and an Expectation Maximisation (EM) algorithm to estimate the latent cluster means.  It also employs Monte Carlo EM for categorical data.  \textit{clustMD} supports mixed data so is appropriate in our context.  To use \textit{clustMD} the data must be reordered to have continuous variables first, followed by ordinal variables and then nominal variables.  For our data, the computational runtime was about 50\% of \textit{JAGS} MCMC, approximately 62 hours.

\section{Results}\label{sec2}

A Dell XPS 7590 laptop with an 8 CPU Intel core i9 processor and 64GB memory was used for this work.  This computer also has an NVIDIA GeForce GTX 1650 GPU but it was not used.

\subsection{Logistic Regression Model}

\begin{figure}[hbt!]
\centering
\includegraphics[width=0.6\linewidth]{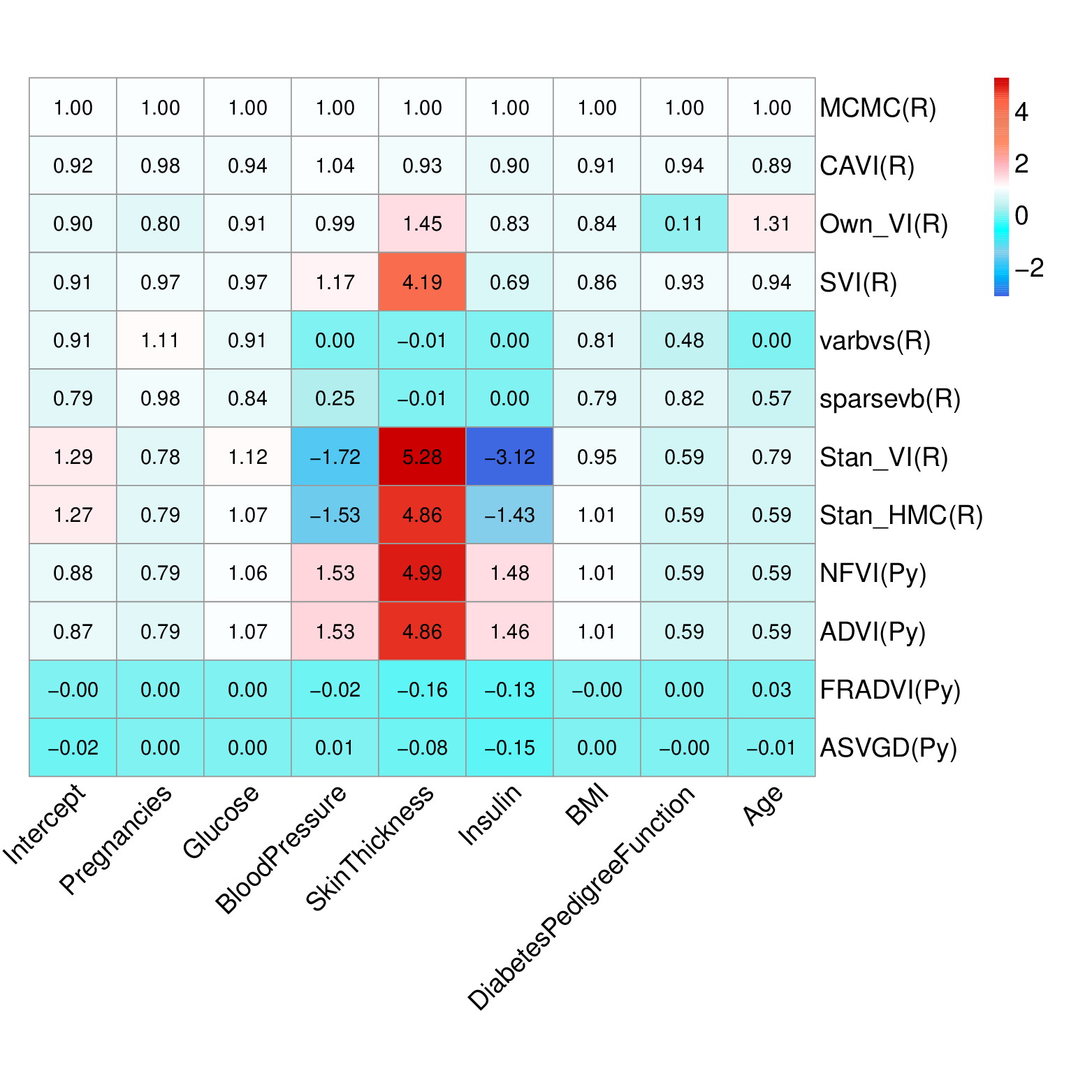}
\captionof{figure}{Average coefficient mean over the 5 folds normalized to the MCMC model (top row). The cell values are calculated by simply normalizing the true values to the corresponding MCMC true value i.e. $cell = \frac{\beta_{true}}{\beta_{mcmc}}$. The programming environment is indicated in the y-axis labels by R for R programming and Py for Python programming. All standard errors are less than 0.25 of the mean SE for MCMC.}
\label{fig:coeffs}
\end{figure}

\begin{figure}[hbt!]
\centering
\includegraphics[width=0.8\linewidth]{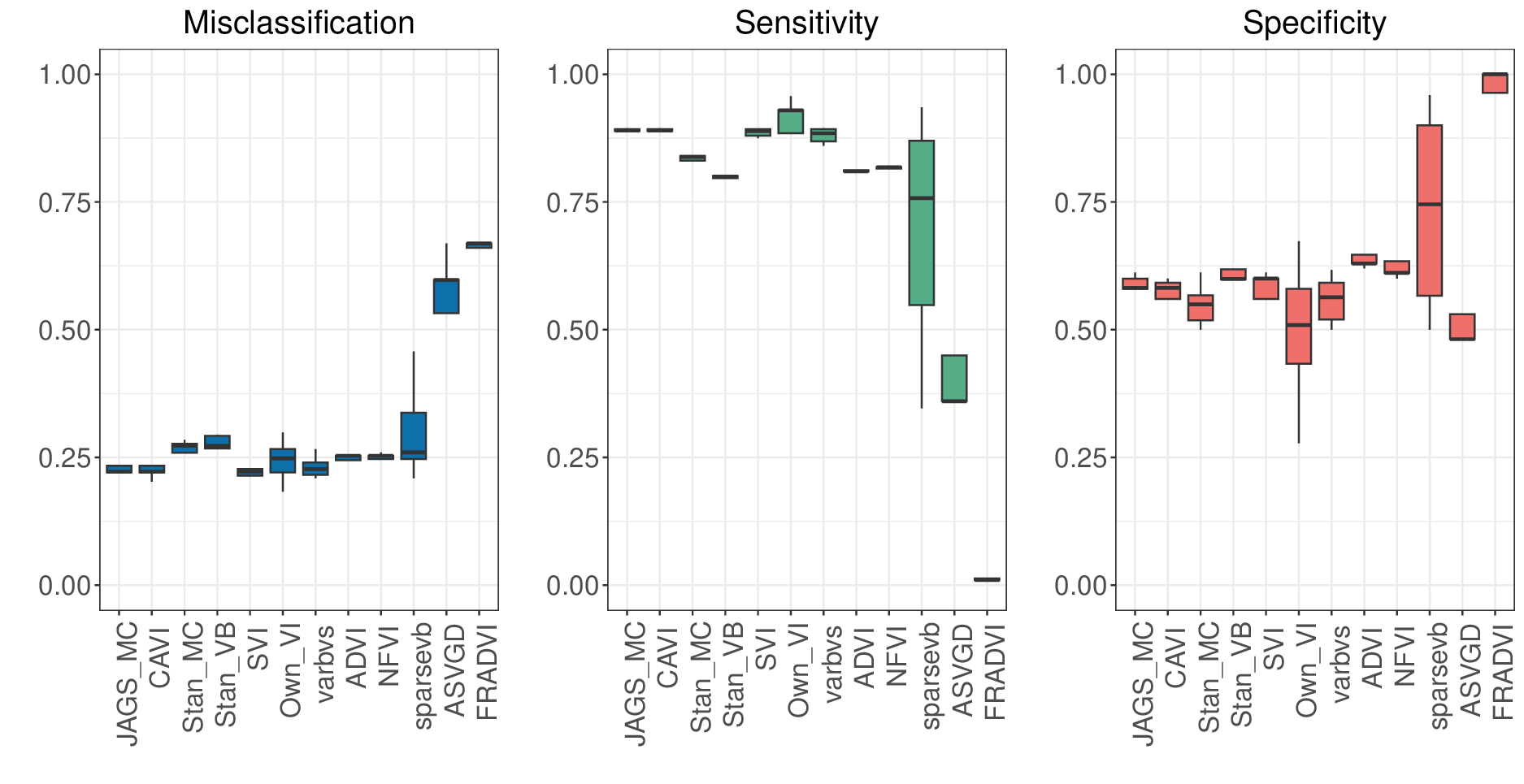}
\captionof{figure}{Predictive performance for 5-fold cross validation. MCMC is the baseline.  The purpose of running cross validation is to check the model stability across data slices.}
\label{fig:perf}
\end{figure}

\begin{figure}[hbt!]
\centering
\includegraphics[width=0.6\linewidth]{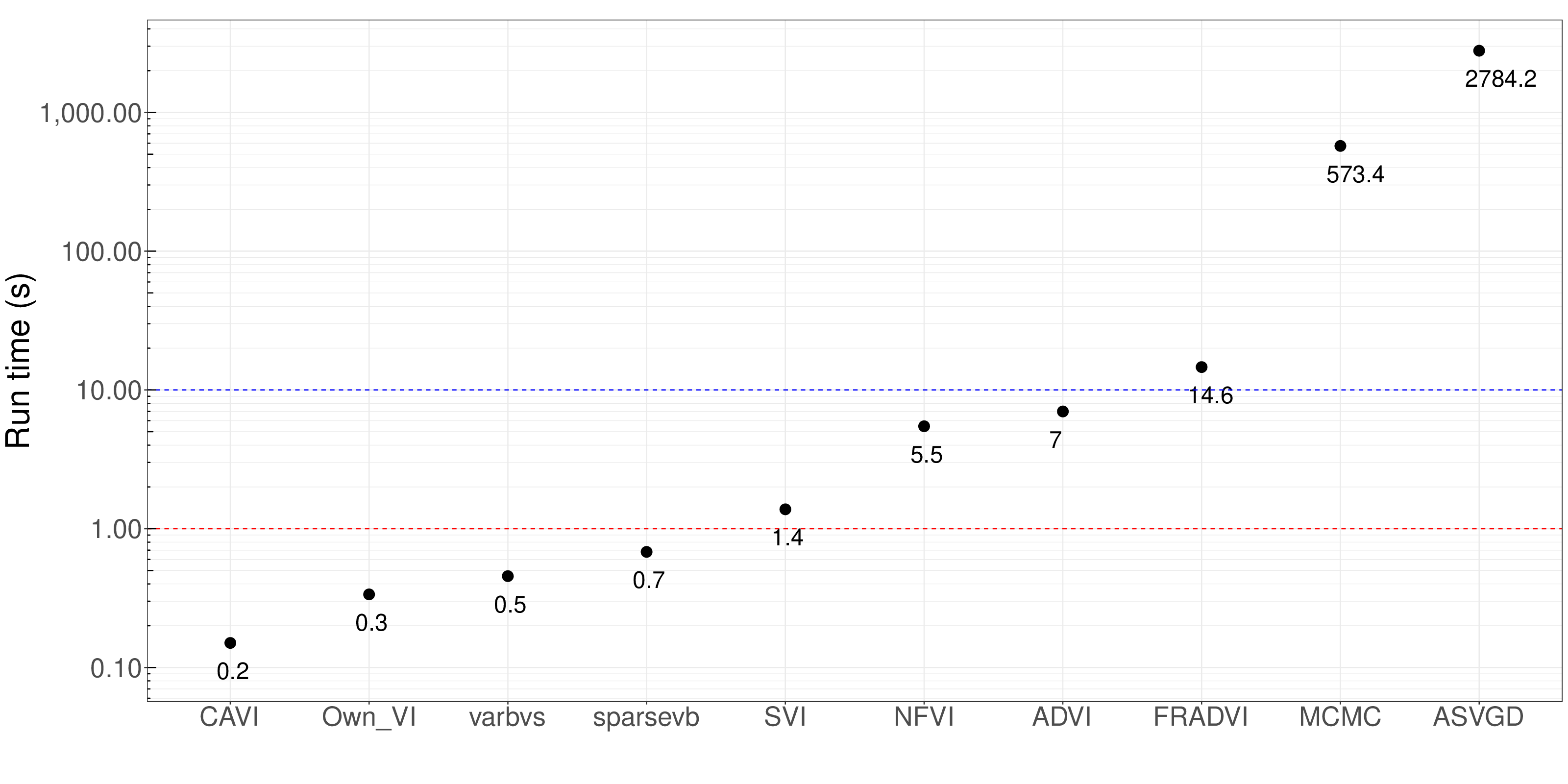}
\captionof{figure}{Computational run times in seconds. Four methods produced sub-second performance (red dotted line) and three had $<$10 seconds run time (blue dotted line). These are very significant improvements compared with MCMC ($\sim$10 minutes).  All results are using the Dell XPS 7590 laptop mentioned at the start of this section.}
\label{fig:runtime}
\end{figure}

Eleven VB implementations were compared against several MCMC implementations.  We used \textit{JAGS} MCMC as our baseline comparison. The mean of the model coefficients averaged over the 5 folds was compared with the MCMC baseline (Figure~\ref{fig:coeffs}).  The heatmap indicates some variables are more challenging across multiple VB methods, e.g. Skin Thickness and Insulin, which have a high proportion of zero values (Table~\ref{table:allResults4}). CAVI, however, returns coefficients very close to MCMC and it returns comparable empirical performance (Figure~\ref{fig:perf}) and computational performance (Figure~\ref{fig:runtime}). Overall the closed-form conditionally conjugate VB methods outperform automatic black-box VB implementations both empirically and computationally. Two Python implementations from the package \textit{PyMC3}, Full-rank ADVI (FRADVI) and Amortized Stein Variational Gradient Descent (ASVGD) could not be configured to produce reasonable results despite a full grid search across their algorithm hyperparameters.

%\begin{figure}[hbt!]
%\centering
%\includegraphics[width=0.6\linewidth]{figs/Figure3.pdf}
%\label{fig:coeffs}
%\captionof{figure}{Coefficient mean as a proportion of MCMC model. The programming environment is indicated in the y-axis labels by R for R programming and Py for Python programming}
%\label{fig:coeffs}
%\end{figure}

It is notable from Figure~\ref{fig:coeffs} that the sign for some coefficients is different to that of MCMC, especially the coefficients for Blood Pressure and Insulin for both \textit{Stan} implementations.  The results data in Table ~\ref{table:allResults3} shows that \textit{JAGS} MCMC returns negative coefficients for both variables whereas \textit{Stan} HMC and VB return positive coefficients for these variables (apart from one \textit{Stan} VB fold). We used flat priors for this model so perhaps more informative priors are required for the \textit{Stan} methods, as mentioned in Chapter 1 of the Stan user's guide. These implementations might also be more influenced by data containing a high proportion of zeros.

The empirical predictive performances closest to MCMC were CAVI, SVI~\citep{durante2019conditionally}, and \textit{varbvs}~\citep{carbonetto2017varbvs} (Figure~\ref{fig:perf}).  All three are mean-field methods that require analytical derivation of the optimization ELBO in contrast to automatic methods ADVI and NFVI~\citep{salvatier2016probabilistic}. All five methods remained reasonably stable across the five cross-validation folds.

%\begin{figure}[hbt!]
%\centering
%\includegraphics[width=0.6\linewidth]{figs/Figure4.pdf}
%\captionof{figure}{Predictive performance. MCMC is the baseline and was validated against the R \textit{glm} function}
%\label{fig:perf}
%\end{figure}
%
%\begin{figure}[hbt!]
%\centering
%\includegraphics[width=0.6\linewidth]{figs/Figure5.pdf}
%\captionof{figure}{Computational run times in seconds. Four methods produced sub-second performance (red dotted line) and three had $<$10 seconds run time (blue dotted line). These are very significant improvements compared with MCMC ($\sim$10 minutes).  All results are using the Dell XPS 7590 laptop mentioned at the start of this section.}
%\label{fig:runtime}
%\end{figure}

\subsection{Composite Bayes LCA/Regression Model}

The baseline results using the Northeast region subset of Optum\textsuperscript{\texttrademark} obtained from \textit{JAGS} were close to the Hubbard et al. results (Table ~\ref{table:comp}).  The differences in biomarker shift can be explained by the wider patient geographic catchment and differences in missing data for PEDSnet and Optum\textsuperscript{\texttrademark}. The striking difference for Endocrinologist visit sensitivity could be due to the much smaller proportion of such visits occurring in the Optum\textsuperscript{\texttrademark} data i.e. 63.43\% in PEDSnet EHR versus 5.9\% in Optum\textsuperscript{\texttrademark} EHR. Following these results, we are confident in the extensibility of the model to the same disease area with different EHR data.  Unfortunately, the computational performance is poor using \textit{JAGS}.  For example, a subset of $\sim$38,000 observations taking 16 hours to run.

\begin{center}
\captionof{table}{Comparison of Optum\textsuperscript{\texttrademark} results with Hubbard et al. PEDSnet using the same JAGS LCA model published in Hubbard et al. GitHub~\citep{hubbard2019bayesian})}
\vspace{3mm}
\scalebox{0.80}{
\begin{tabular}{llll}
\toprule
\textbf{}                               & \multicolumn{2}{l}{\textbf{Posterior Mean (95\% CI)}}                   \\
                                        & (a) PEDSnet data                   & (b) Optum\textsuperscript{\texttrademark} data \\
\midrule
                                        &                                    &                                    \\
T2DM code sensitivity (expit($\beta^W_{10} + \beta^W_{11}$))                   & 0.17 (0.15, 0.20)       &  0.15 (0.12, 0.18)      \\[1mm]
T2DM code specificity (1-expit($\beta^W_{10}$))                   & 1.00 (1.00, 1.00)       &  1.00 (1.00, 1.00)                   \\[1mm]
Endocrinologist visit code sensitivity (expit($\beta^W_{20} + \beta^W_{21}$))  & 0.94 (0.92, 0.95)       &  0.18 (0.15, 0.21)      \\[1mm]
Endocrinologist visit code specificity (1-expit($\beta^W_{20}$))  & 0.93 (0.93, 0.94)       &  0.99 (0.98, 0.99)                   \\[1mm]
Metformin code sensitivity (expit($\beta^P_{10} + \beta^P_{11}$))              & 0.31 (0.28, 0.35)       &  0.40 (0.36, 0.44)      \\[1mm]
Metformin code specificity (1-expit($\beta^P_{10}$))              & 0.99 (0.99, 0.99)       &  0.98 (0.98, 0.99)                   \\[1mm]
Insulin code sensitivity (expit($\beta^P_{20} + \beta^P_{21}$))                & 0.66 (0.61, 0.70)       &  0.55 (0.51, 0.59)      \\[1mm]
Insulin code specificity (1-expit($\beta^P_{20}$))                & 1.00 (1.00, 1.00)       &  1.00 (1.00, 1.00)                   \\[1mm]
Mean shift in HbA1c ($\beta^Y_{12}$)                     & 3.15 (3.06, 3.24)       &  4.80 (4.72, 4.81)                            \\[1mm]
Mean shift in Glucose ($\beta^Y_{11}$)                   &            90.62 (90.25, 91.00)    &            89.30 (89.10, 90.01)    \\[1mm]
\bottomrule
\label{table:comp}
\end{tabular}}
\end{center}

We tested posterior diagnostics, goodness of fit diagnostics and the empirical performance of the composite model.  \textit{Stan} has comprehensive posterior diagnostics available via the \textit{posterior}~\citep{rposterior2023} and \textit{bayesplot}~\citep{gabry2017bayesplot} R packages.  The \textit{loo} R package \cite{vehtari2017practical} provides goodness of fit diagnostics based on Pareto Smoothed Importance Sampling (PSIS)~\citep{vehtari2015pareto}, leave-one-out cross-validation and the Watanabe-Akaike/Widely Applicable information criterion (WAIC)~\citep{magnusson2020leave, vehtari2017practical}.

\subsubsection{Posterior Diagnostics}

\begin{figure}[hbt!]
\centering
  \includegraphics[width=0.8\linewidth]{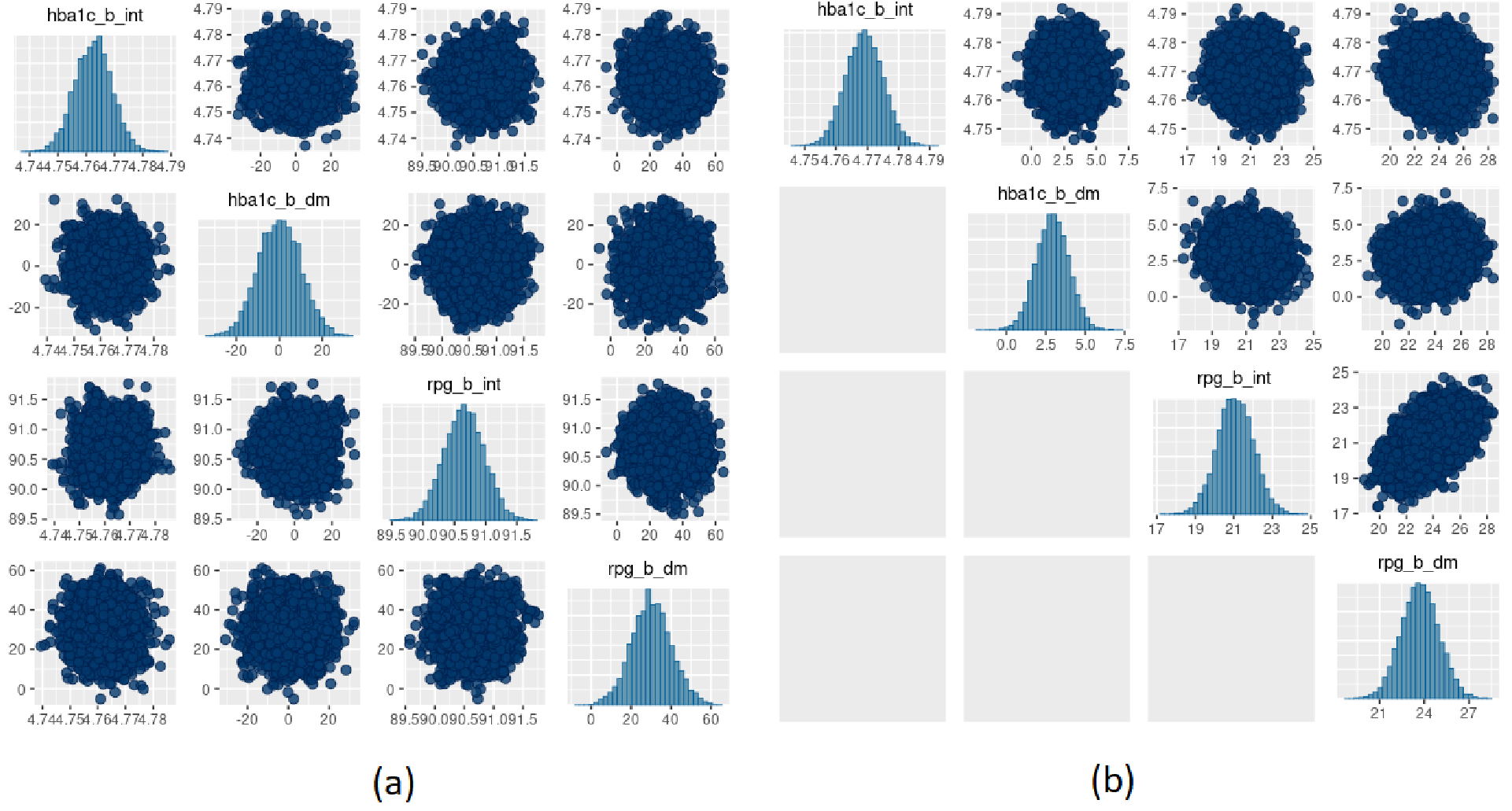}
\caption{\textit{bayesplot} pairs plots for (a) \textit{Stan} HMC and (b) \textit{Stan} VB for the two biomarkers (HbA1c and random plasma glucose, RPG).  Each biomarker contains two priors as defined in the model specification (Table ~\ref{table:modelSpec}).  The b\_int prior is the multivariate normal, $\beta^Y_j$, representing a biomarker test result for normal, non-diabetes patients and the b\_dm prior encodes known information on predictive accuracy containing values corresponding to a ROC AUC of 0.95. as described in Hubbard et al. Section 2.2 and 2.5~\citep{hubbard2019bayesian}.}
\label{fig:bayesplot}
\end{figure}

The posterior diagnostics plots for the biomarkers in Figure~\ref{fig:bayesplot} show that, for (a) MCMC, we see no evidence of collinearity and the posterior means appear close to those we obtained with \textit{JAGS} (b).  The \textit{mcmc pairs} plot for VB appears reasonable for HbA1c but not for the Glucose biomarker \textit{rpg\_b\_int} prior.  There appears to be a correlation between the glucose priors and the expected value for Glucose is far from that obtained with MCMC (Table \ref{table:allResults}). We were unable to fully explain this correlation and the Glucose mode far from that reported by MCMC. We made several amendments to the model specification along with algorithm hyperparameter tuning that all returned the same effect for Glucose. Since there is a single chain for VB there is only the top triangular set, which represents 100\% of the posterior samples.  This type of plot does not communicate the relative variances of the posteriors.

\clearpage
\begin{center}
\captionof{table}{Comparison of composite LCA/regression model results for clinical attributes}
\vspace{3mm}
\scalebox{0.80}{
\begin{tabular}{lllll}
\toprule
\textbf{}                               & \multicolumn{3}{l}{\textbf{Posterior Mean (95\% CI)}}                                                     \\
                                        & (a) JAGS Gibbs MCMC           & (b) Stan HMC           & (c) Stan VB                                      \\
\midrule
                                        &                         &                       &                                                         \\
T2DM code sensitivity (expit($\beta^W_{10} + \beta^W_{11}$))                   &  0.15 (0.12, 0.18)      &  0.10 (0.09, 0.11)    & 0.12 (0.10, 0.12)      \\[1mm]
T2DM code specificity (1-expit($\beta^W_{10}$))                     &  1.00 (1.00, 1.00)      &  1.00 (0.99, 1.00)    & 0.99 (0.99, 0.99)      \\[1mm]
Endocrinologist visit code sensitivity (expit($\beta^W_{20} + \beta^W_{21}$))   &  0.18 (0.15, 0.21)      &  0.20 (0.18, 0.21)    & 0.22 (0.19, 0.22)      \\[1mm]
Endocrinologist visit code specificity (1-expit($\beta^W_{20}$))  &  0.99 (0.98, 0.99)      &  0.98 (0.97, 0.99)    & 0.97 (0.97, 0.99)      \\[1mm]
Metformin code sensitivity (expit($\beta^P_{10} + \beta^P_{11}$))              &  0.40 (0.36, 0.44)      &  0.21 (0.20, 0.21)    & 0.19 (0.19, 0.20)      \\[1mm]
Metformin code specificity (1-expit($\beta^P_{10}$))               &  0.98 (0.98, 0.99)      &  0.93 (0.92, 0.93)    & 0.93 (0.92, 0.94)      \\[1mm]
Insulin code sensitivity (expit($\beta^P_{20} + \beta^P_{21}$))                &  0.55 (0.51, 0.59)      &  0.35 (0.31, 0.35)    & 0.20 (0.19, 0.20)      \\[1mm]
Insulin code specificity (1-expit($\beta^P_{20}$))                &  1.00 (1.00, 1.00)      &  1.00 (0.99, 1.00)    & 1.00 (0.99, 1.00)      \\[1mm]
Mean shift in HbA1c ($\beta^Y_{12}$)                     &  4.80 (4.72, 4.81)      &  4.77 (4.76, 4.78)    & 4.77 (4.75, 4.78)      \\[1mm]
Mean shift in glucose ($\beta^Y_{11}$)                   & 89.30 (89.10, 90.01)                                     & 88.59 (88.48, 88.71)                                   & 22.80 (21.06, 24.92) \\
\bottomrule
\label{table:allResults}
\end{tabular}}
\end{center}

%\begin{figure}[hbt!]
%\centering
%  \includegraphics[width=0.8\linewidth]{figs/Figure6.pdf}
%\caption{\textit{bayesplot} pairs plots for (a) \textit{Stan} HMC and (b) \textit{Stan} VB for the two biomarkers.  Each biomarker contains two model priors as defined in the model specification (Table ~\ref{table:modelSpec}).  The b\_int prior is the multivariate normal, $\beta^Y_j$ and the b\_dm prior encodes known information on predictive accuracy containing values corresponding to a ROC AUC of 0.95. as described in Hubbard et al. Section 2.2 and 2.5~\citep{hubbard2019bayesian}.}
%\label{fig:bayesplot}
%\end{figure}

\subsubsection{Goodness of fit}

\begin{figure}[hbt!]
\centering
  \includegraphics[width=0.8\linewidth]{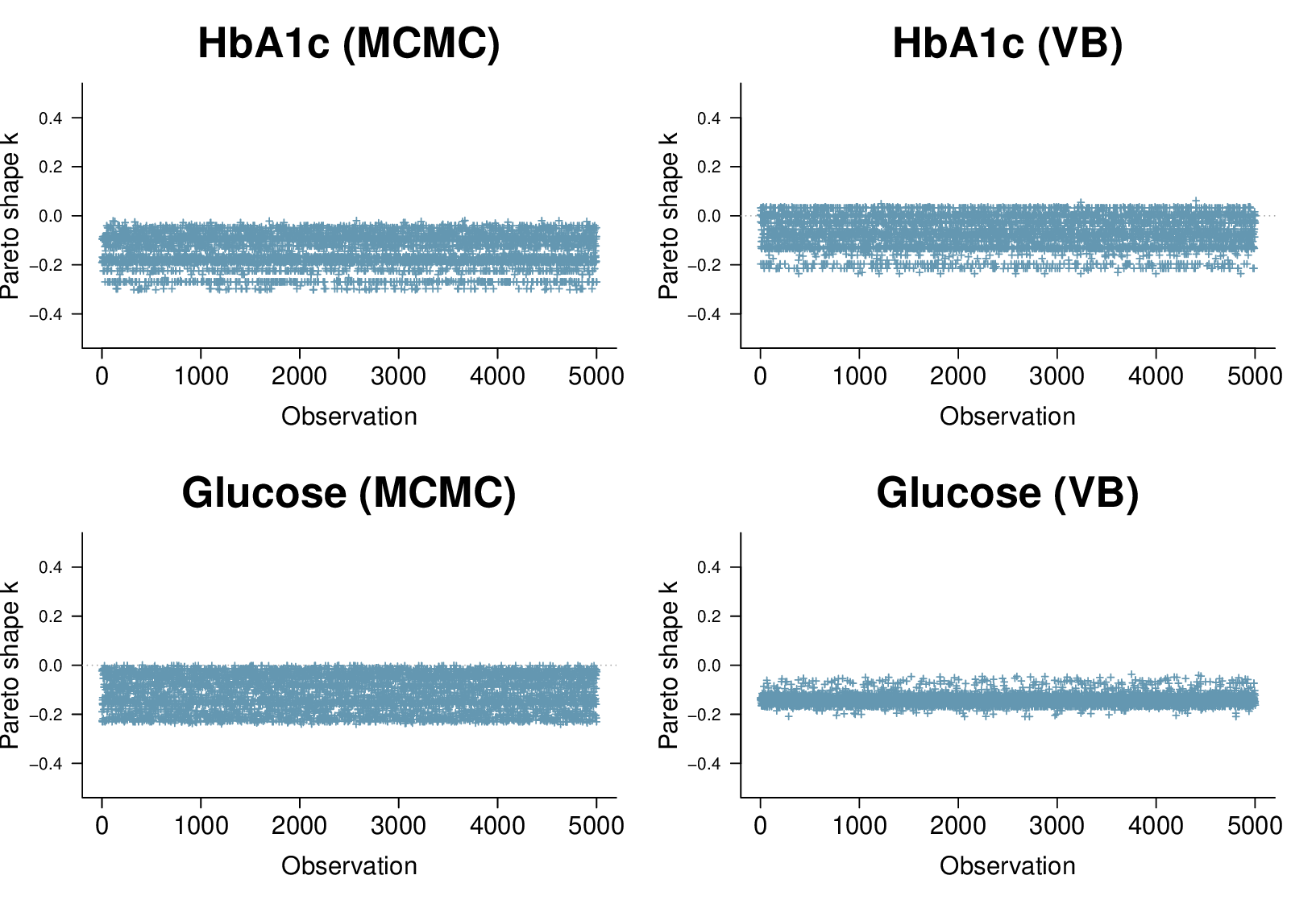}
  \caption{PSIS plots for the two biomarkers, HbA1c (top row) and Random Glucose (bottom row) under MCMC (left) and VB (right). Both biomarkers are well below $k$=0.5.  The HbA1c biomarker is slightly worse for VB with a segment of observations above 0 but is well within good territory.  The Random Glucose biomarker appears better in VB compared to MCMC but we know that the expected value obtained by VB for Glucose is not as close to the true value obtained by MCMC.}
  \label{fig:loo}
\end{figure}

We used approximate leave-one-out cross-validation from the R \textit{loo} package to evaluate the goodness of fit for the model.  \textit{loo} uses log-likelihood point estimates from the model to measure its predictive accuracy against training samples generated by Pareto Smoothed Importance Sampling (PSIS)~\citep{vehtari2015pareto}.  The PSIS shape parameter $k$ is used to assess the reliability of the estimate.  If $k <$ 0.5 the variance of the importance ratios is finite and the central limit theorem holds.  If $k$ is between 0.5 and 1 the variance of the importance ratios is infinite but the mean exists.  If $k >$ 1 the variance and mean of the importance ratios do not exist. The results for the two biomarkers (Figure~\ref{fig:loo}) show all of the n=38,000 observations are in a $k <$ 0.5 range.  It appears VB performs approximately as well as MCMC in the context of the PSIS metric.

%\begin{figure}[hbt!]
%\centering
%  \includegraphics[width=0.8\linewidth]{figs/Figure7.pdf}
%  \caption{PSIS plots for the two biomarkers, HbA1c (top row) and Random Glucose (bottom row) under MCMC (left) and VB (right). Both biomarkers are well below $k$=0.5.  The HbA1c biomarker is slightly worse for VB with a segment of observations above 0 but is well within good territory.  The Random Glucose biomarker appears better in VB compared to MCMC but we know that the expected value obtained by VB for Glucose is not as close to the true value obtained by MCMC.}
%  \label{fig:loo}
%\end{figure}

\subsubsection{Empirical Performance}

The model sensitivity analysis for the indicator variables shows good agreement with MCMC (Table ~\ref{table:allResults}). The mean shift estimates for biomarkers Glucose variable under ADVI.

\subsubsection{Comparison with Maximum Likelihood Approach}

For \textit{clustMD} we take cluster 1 as the T2DM class as it is the minority cluster.  The cluster means parallel coordinates plot in Figure~\ref{fig:mle}a indicates similar HbA1c (4.81) and Glucose (89.9) levels compared to the Bayesian model. Cluster 1 has a significantly larger variance (Figure~\ref{fig:mle}b) possibly due to the high imbalance of the T2DM positive class in the data.

%\begin{figure}[hbt!]
%\centering
%  \includegraphics[width=0.45\linewidth]{figs/Figure8a.pdf}
%  \includegraphics[width=0.45\linewidth]{figs/Figure8b.pdf}
%\caption{\textit{clustMD} plots running 2 latent clusters.  (a) shows a parallel coordinates plot for all variables, (b) shows the cluster variances for all variables.}
%\label{fig:mle}
%\end{figure}

%\clearpage
\section{Discussion}\label{sec3}

We compared the use of VB and MCMC for Bayesian latent class analysis and logistic regression models with the objective of scaling a composite LCA/regression patient phenotyping model proposed by Hubbard et al.~\citep{hubbard2019bayesian} We compared a number of alternative methods for estimating parameters in the model to identify if the intrinsic computational limitations from MCMC can be overcome in a real-world clinical setting using VB. We compared VB and MCMC for logistic regression and latent class analysis for similar clinical settings with two data sets, Pima Indians' diabetes and Optum\textsuperscript{\texttrademark} EHR diabetes data. We set out some practical issues in using VB compared to MCMC for these methods on these data such as model stability, data size (primarily in terms  of large N) and mixed discrete and continuous data.  Some practical guidance included balancing accuracy and runtime via hyperparameter settings and amelioration of label switching via setting constraints on the priors to ensure close vicinity to the clinically expected solution. For Bayes LCA, we found a lack of closed-form VB implementations currently available so we used black-box automatic approaches.  We find automatic black-box VB methods as implemented both for the baseline Pima Indian logistic regression model and the Optum\textsuperscript{\texttrademark} EHR composite LCA/regression model are complex to configure and very sensitive to model and prior definition, algorithm hyperparameters and choice of gradient optimiser. The composite Bayes LCA/regression model was significantly more challenging to implement but it was possible to achieve reasonable results with careful model specification and hyperparameter tuning. This however, results in an iterative trial-and-error approach that we find can sometimes be more cumbersome than running a multi-chain MCMC. This work has a number of limitations. In the Optum\textsuperscript{\texttrademark} EHR case study we have not fully explained why the VB posterior prior values for random plasma glucose are so different to those obtained via MCMC (both JAGS and Stan).  This result is a potential question for further work.  Further limitations are the exploration of examples where some (but not all) methods run into issues and a discussion as to why this may be e.g. Kucukelbir et. al., allude to potential avenues in section 5 of their ADVI paper~\citep{kucukelbir2017automatic} and Dhaka et. al. list several challenges with black-box approaches to VB ~\citep{dhaka2021challenges}. Further work could also explore how these methods compare for not only large n but also large p, such as for genetic data.  Since data sparsity is an important aspect of clinical data, we feel a comprehensive investigation of this area is warranted.  We also deliberately restricted our comparison to Bayesian methods so that prior information could be incorporated, so we did not include detailed comparisons with machine learning or frequentist methods. The Pima Indians case study demonstrated that the closed-form conditionally conjugate mean-field approach to VB, even though it has many simplifying assumptions, performed best of the VB approaches, excelling both computationally and in achieving posterior accuracy comparable to the MCMC gold standard for that data set.

% For Original Research articles, please note that the Material and Methods section can be placed in any of the following ways: before Results, before Discussion or after Discussion.

\section*{Conflict of Interest Statement}
%All financial, commercial or other relationships that might be perceived by the academic community as representing a potential conflict of interest must be disclosed. If no such relationship exists, authors will be asked to confirm the following statement:

The authors declare that the research was conducted in the absence of any commercial or financial relationships that could be construed as a potential conflict of interest.

\section*{Author Contributions}

All authors contributed to the study conception and/or data interpretation. Computer programming, Data collection and analysis were performed by BB. All authors contributed to the writing and/or editing of the manuscript and approved the final draft

\section*{Funding}

No sources of funding were sought for this work.

\section*{Data Availability Statement}

Data for the patient phenotyping case study was made available through a third-party license from Optum\textsuperscript{\texttrademark} EHR, a commercial data provider in the USA. Further release of the data set is not possible due to a data use agreement. However, realistic simulated EHR data using the Bayesian LCA model priors is available along with the model output data and the R and Python scripts from our GitHub. The simulated EHR data is the same size and has the same statistical characteristics as the original Optum\textsuperscript{\texttrademark} EHR data. The Pima Indian diabetes data set analyzed for this study can be found in the Kaggle website https://www.kaggle.com/datasets/uciml/pima-indians-diabetes-database.
% Please see the availability of data guidelines for more information, at https://www.frontiersin.org/about/author-guidelines#AvailabilityofData

\bibliographystyle{unsrt}
\bibliography{frontiers}

%%% Make sure to upload the bib file along with the tex file and PDF
%%% Please see the test.bib file for some examples of references

\clearpage
\section*{Figure captions}

%%% Please be aware that for original research articles we only permit a combined number of 15 figures and tables, one figure with multiple subfigures will count as only one figure.
%%% Use this if adding the figures directly in the mansucript, if so, please remember to also upload the files when submitting your article
%%% There is no need for adding the file termination, as long as you indicate where the file is saved. In the examples below the files (logo1.eps and logos.eps) are in the Frontiers LaTeX folder
%%% If using *.tif files convert them to .jpg or .png
%%%  NB logo1.eps is required in the path in order to correctly compile front page header %%%

\begin{figure}[hbt!]
\centering
  \includegraphics[width=0.45\linewidth]{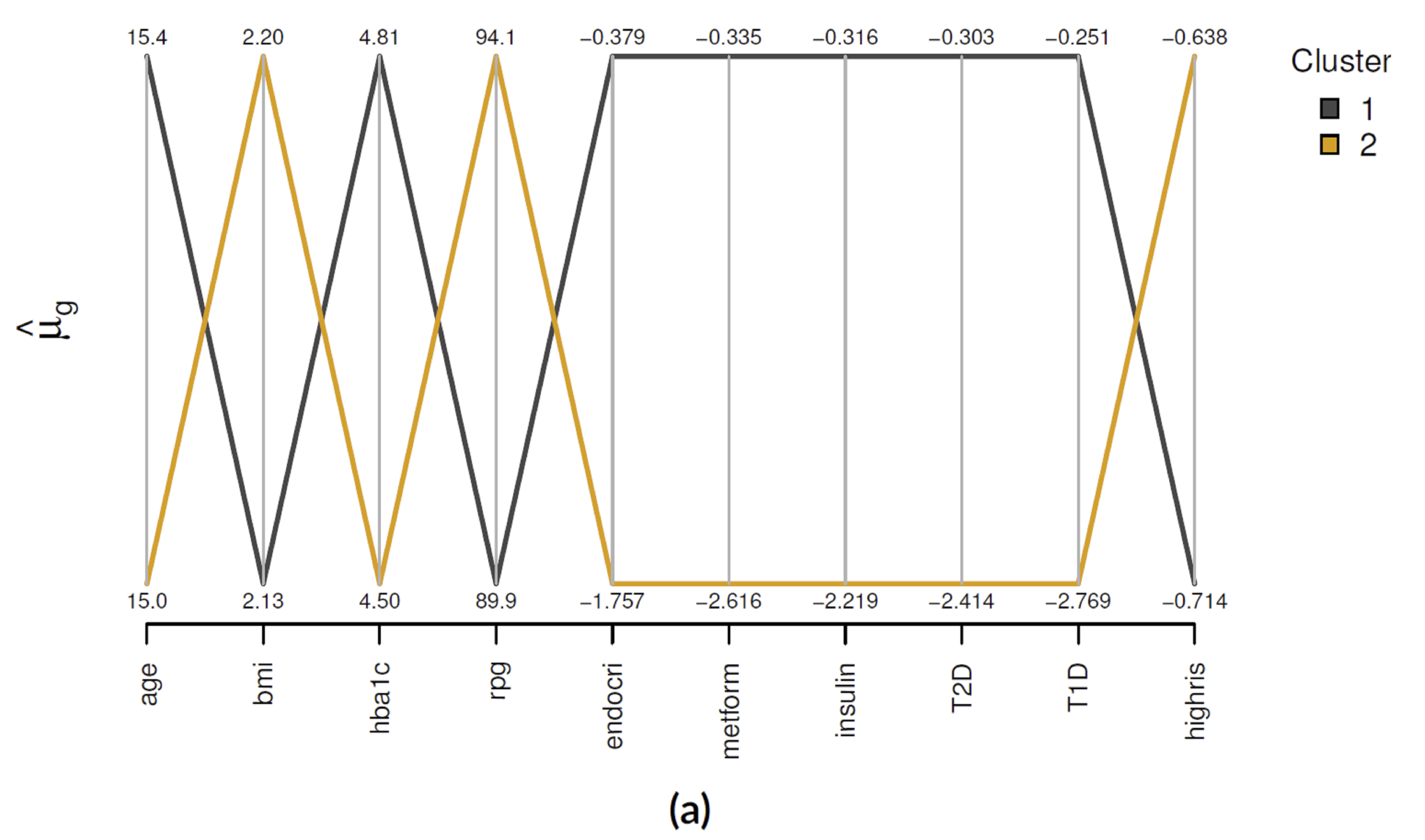}
  \includegraphics[width=0.45\linewidth]{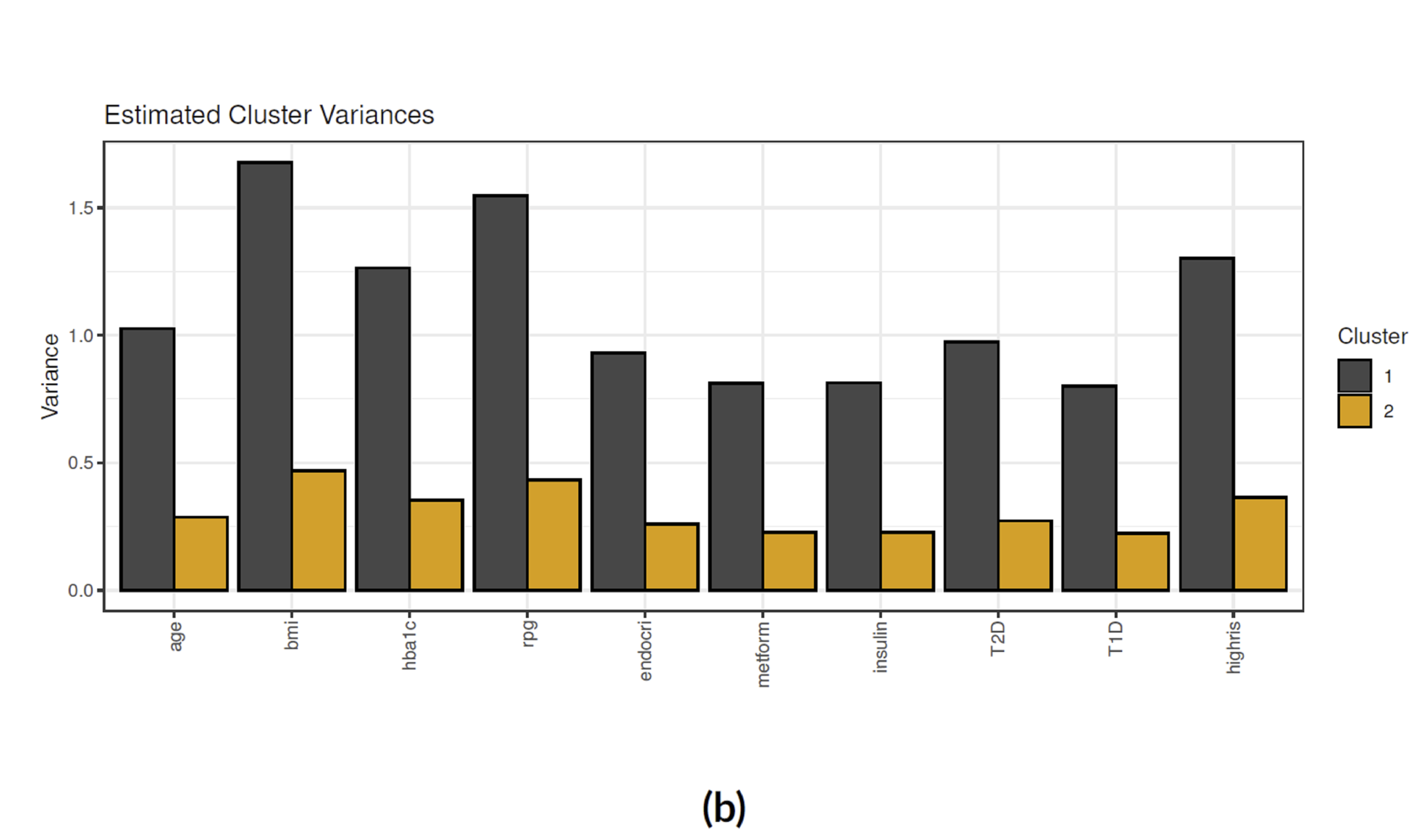}
\caption{\textit{clustMD} plots running 2 latent clusters.  (a) shows a parallel coordinates plot for all variables, (b) shows the cluster variances for all variables.}
\label{fig:mle}
\end{figure}

\begin{figure}[hbt!]
\centering
  \setlength{\fboxsep}{0pt}%
  \setlength{\fboxrule}{0pt}%
  \fbox{\includegraphics[width=0.45\linewidth]{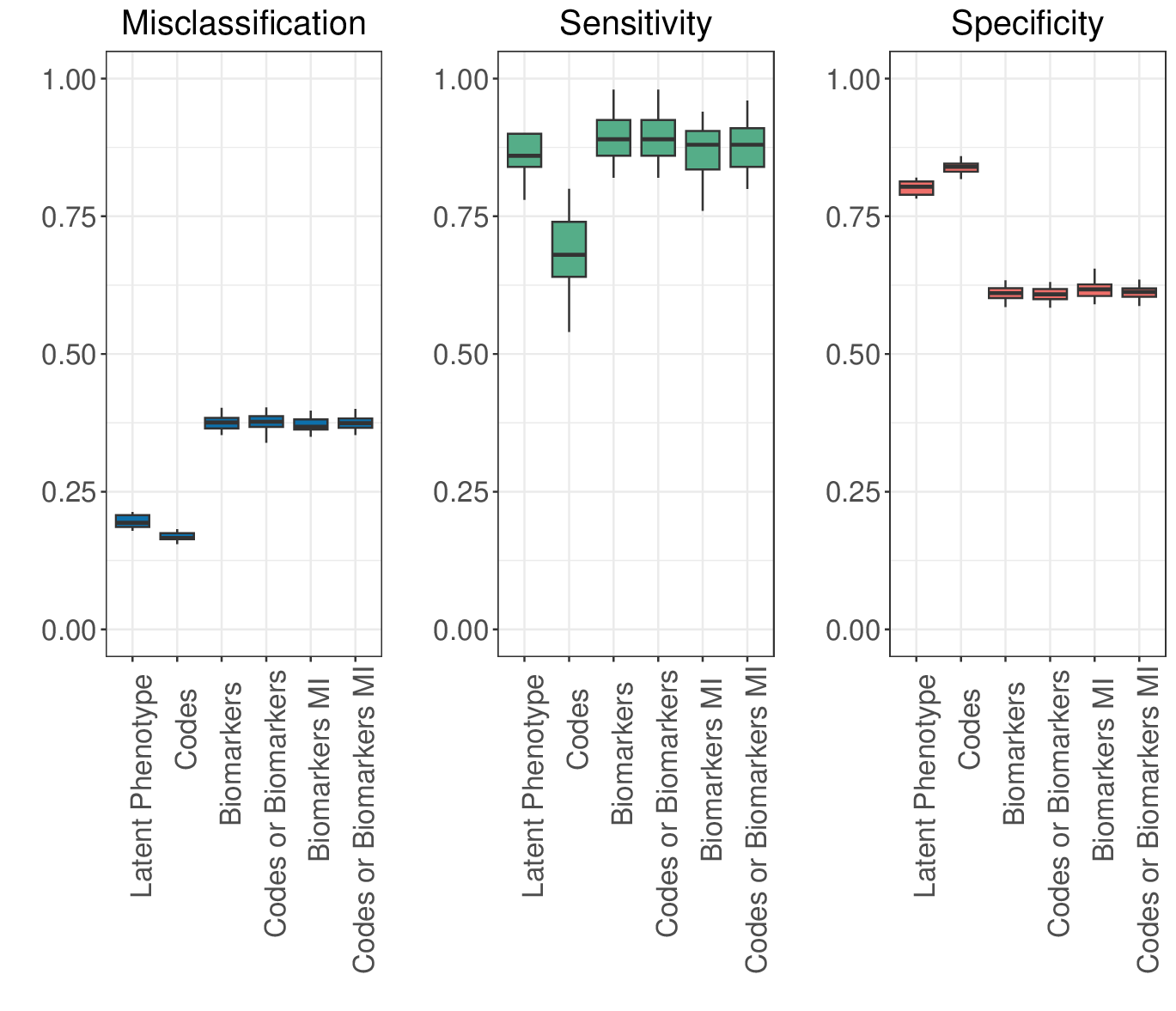}}
  \fbox{\includegraphics[width=0.45\linewidth]{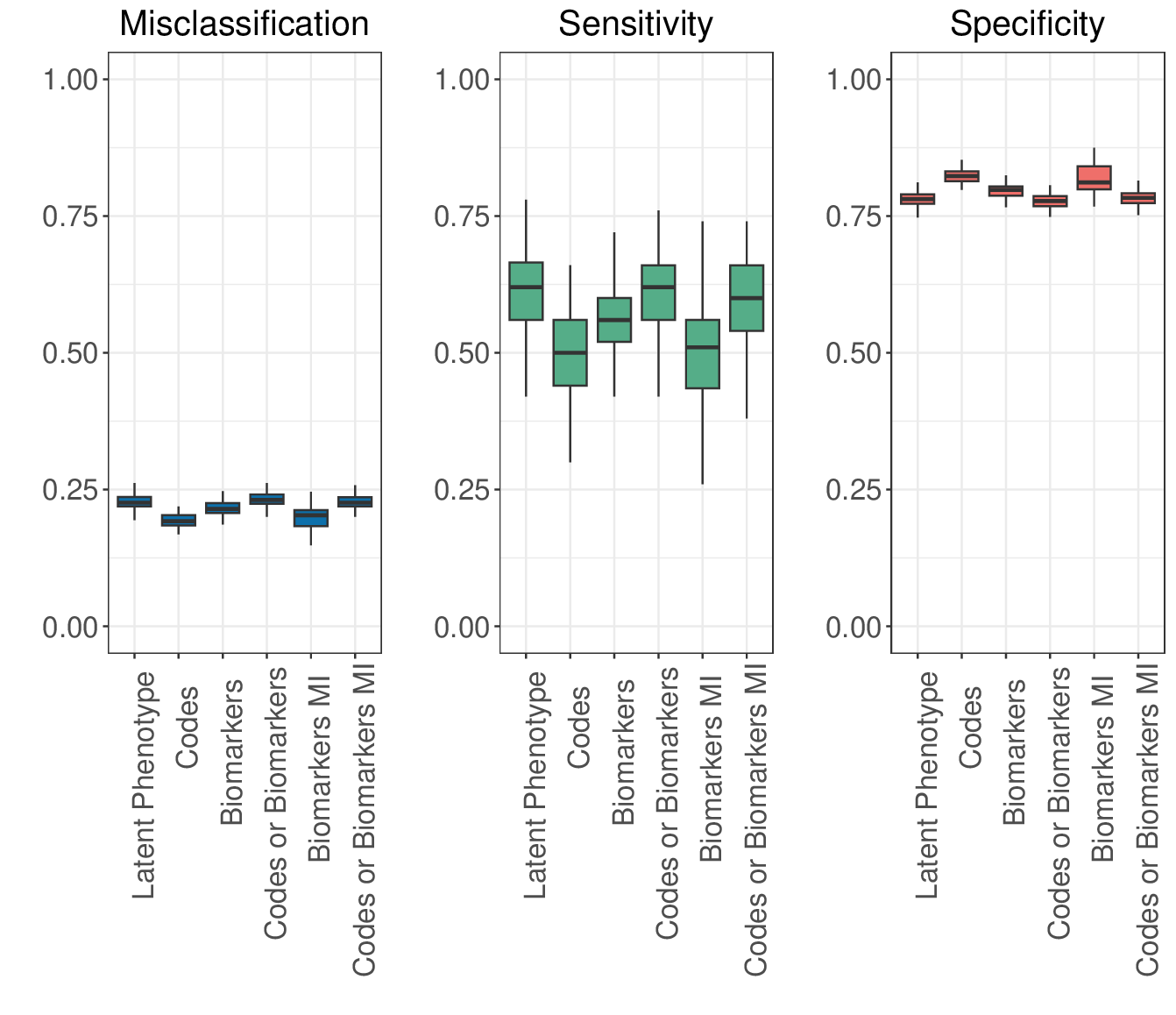}} \\
  \fbox{\includegraphics[width=0.45\linewidth]{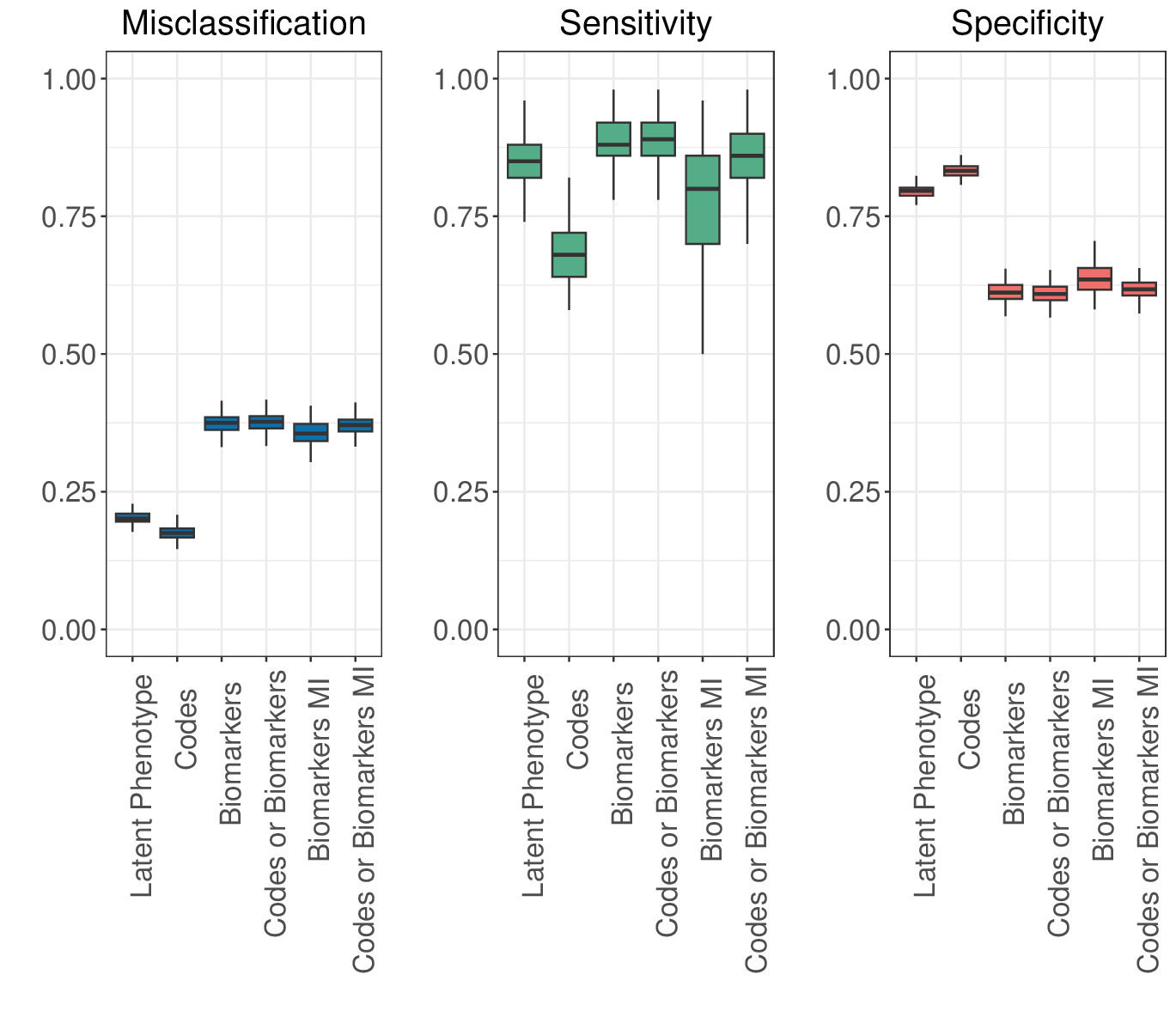}}
  \fbox{\includegraphics[width=0.45\linewidth]{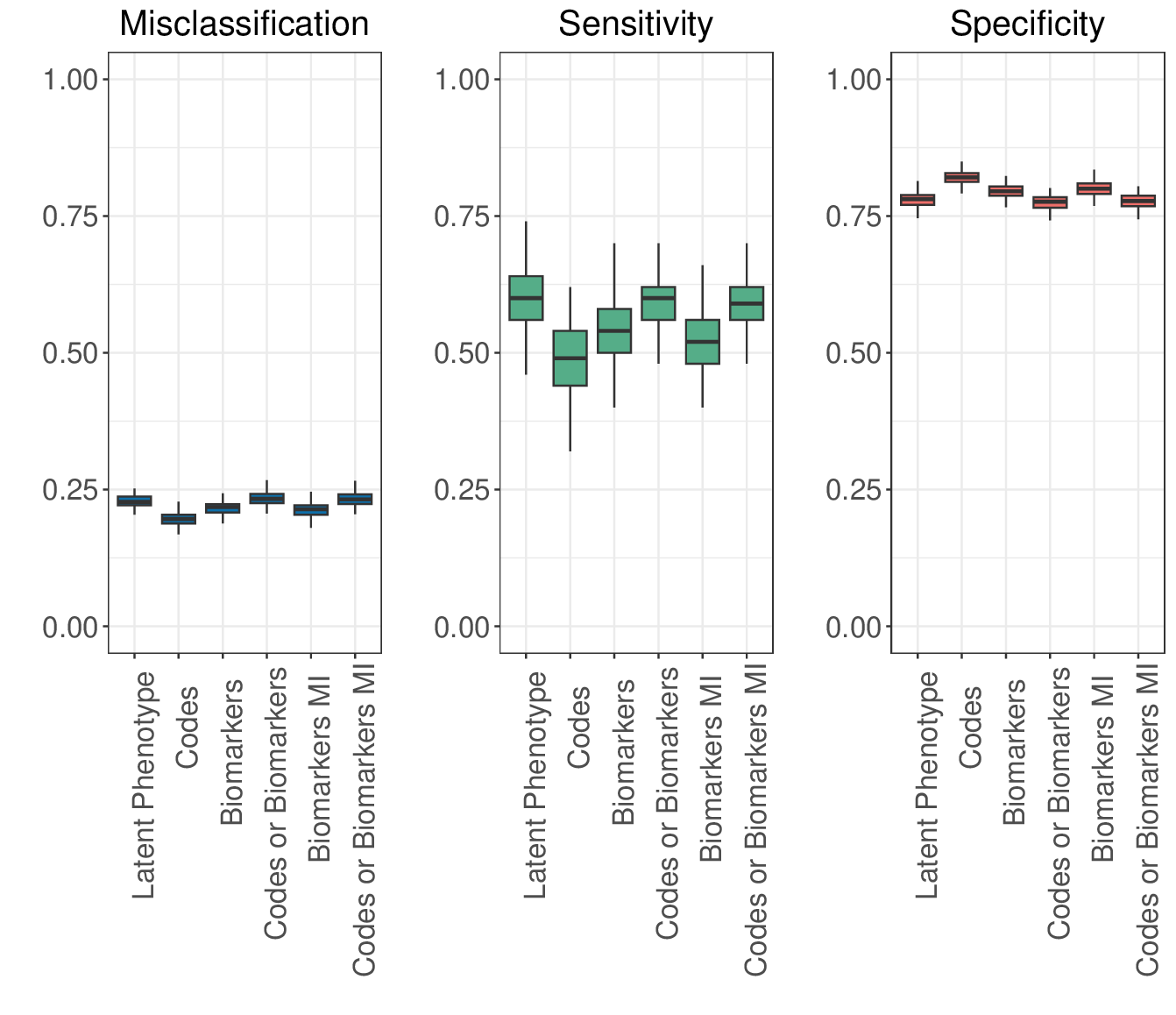}}
\caption{Simulation sensitivity study results comparing MCMC (left column) with VB (right column) for high MAR (top) and low MAR (bottom).}
\label{fig:mle}
\end{figure}

\begin{figure}[hbt!]
\centering
  \setlength{\fboxsep}{0pt}%
  \setlength{\fboxrule}{0pt}%
  \fbox{\includegraphics[width=0.45\linewidth]{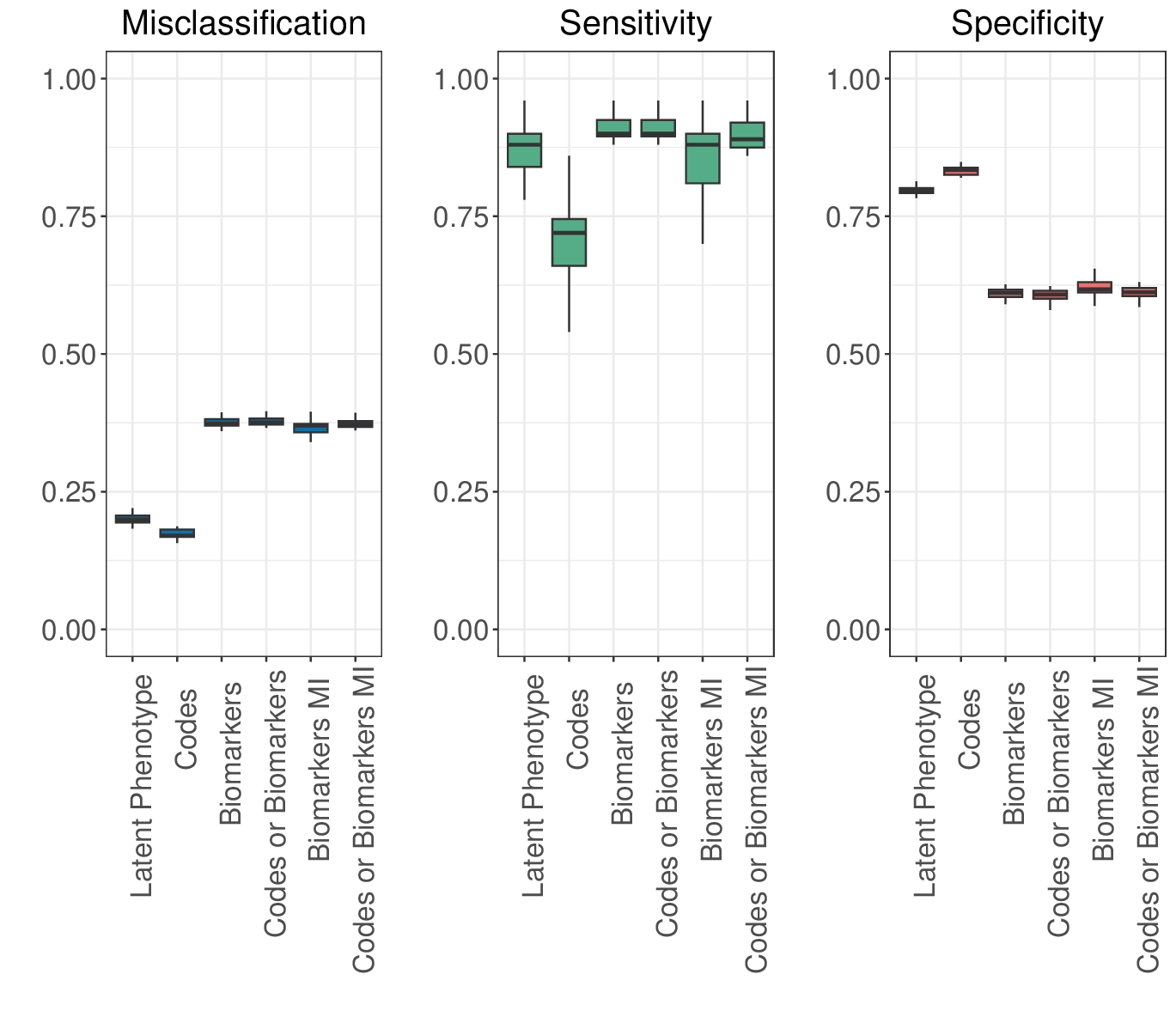}}
  \fbox{\includegraphics[width=0.45\linewidth]{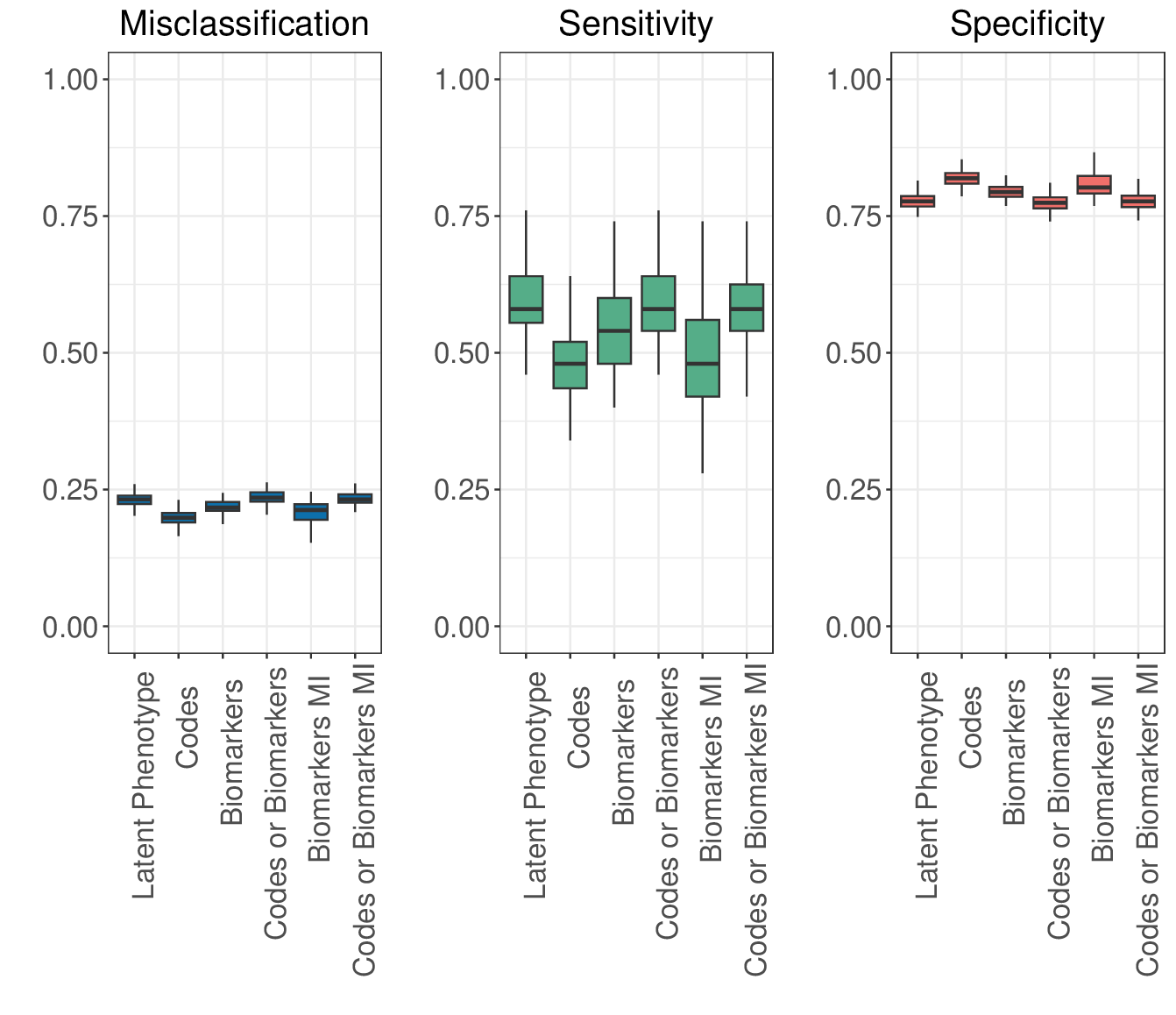}} \\
  \fbox{\includegraphics[width=0.45\linewidth]{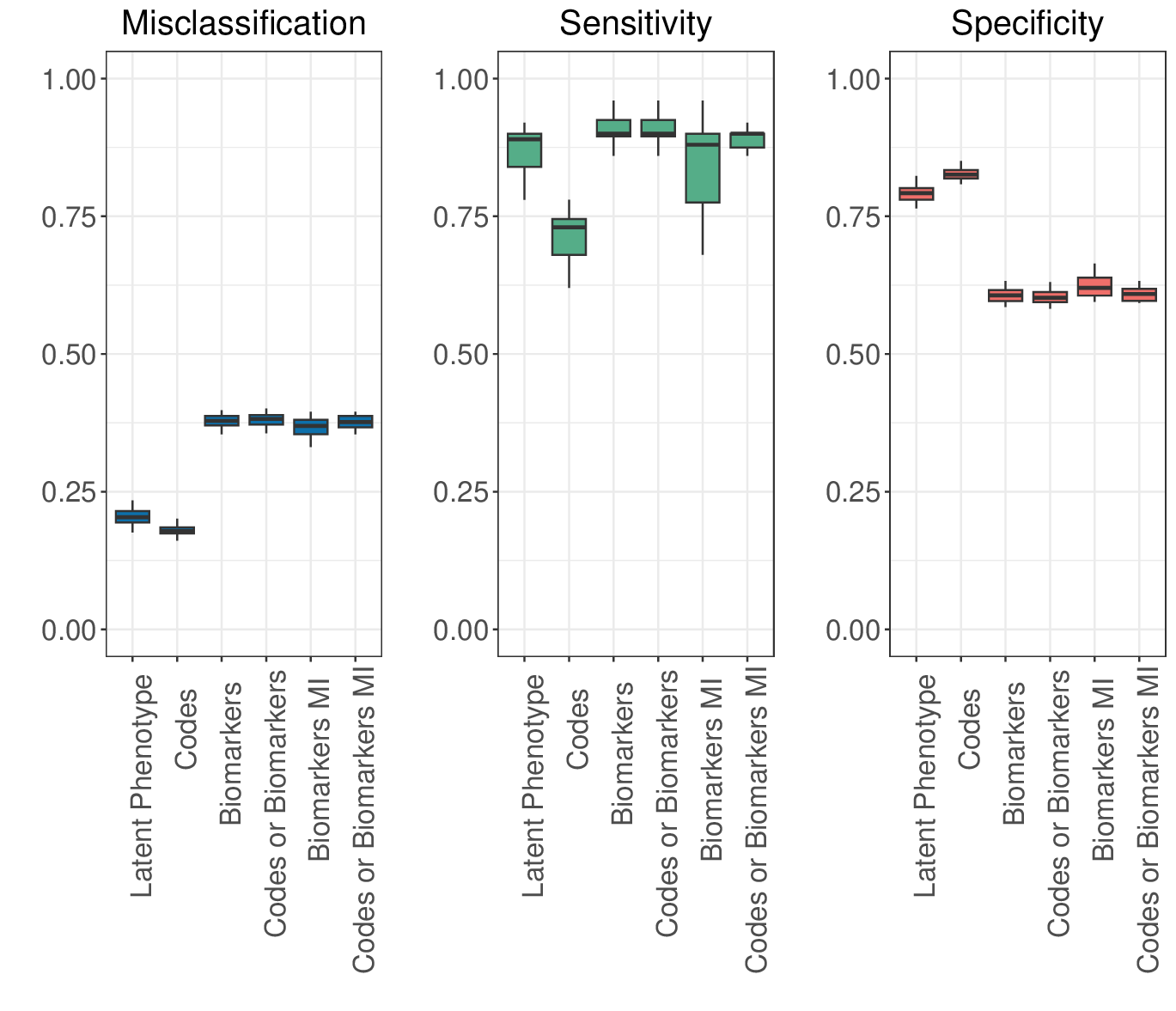}}
  \fbox{\includegraphics[width=0.45\linewidth]{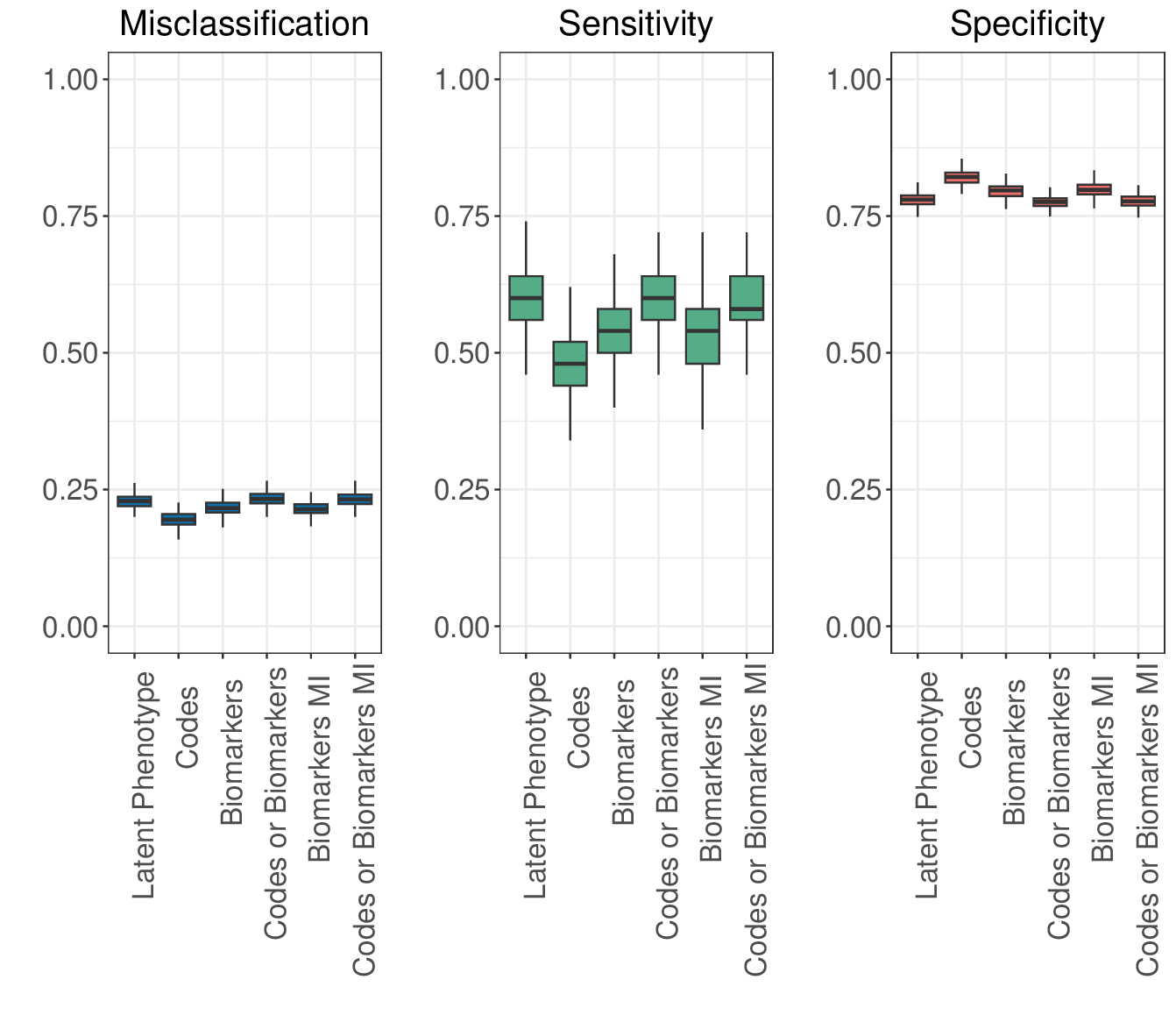}}
\caption{Simulation sensitivity study results comparing MCMC (left column) with VB (right column) for high MNAR (top) and low MNAR (bottom).}
\label{fig:mle}
\end{figure}

\begin{figure}[hbt!]
\centering
  \setlength{\fboxsep}{0pt}%
  \setlength{\fboxrule}{0pt}%
  \fbox{\includegraphics[width=0.45\linewidth]{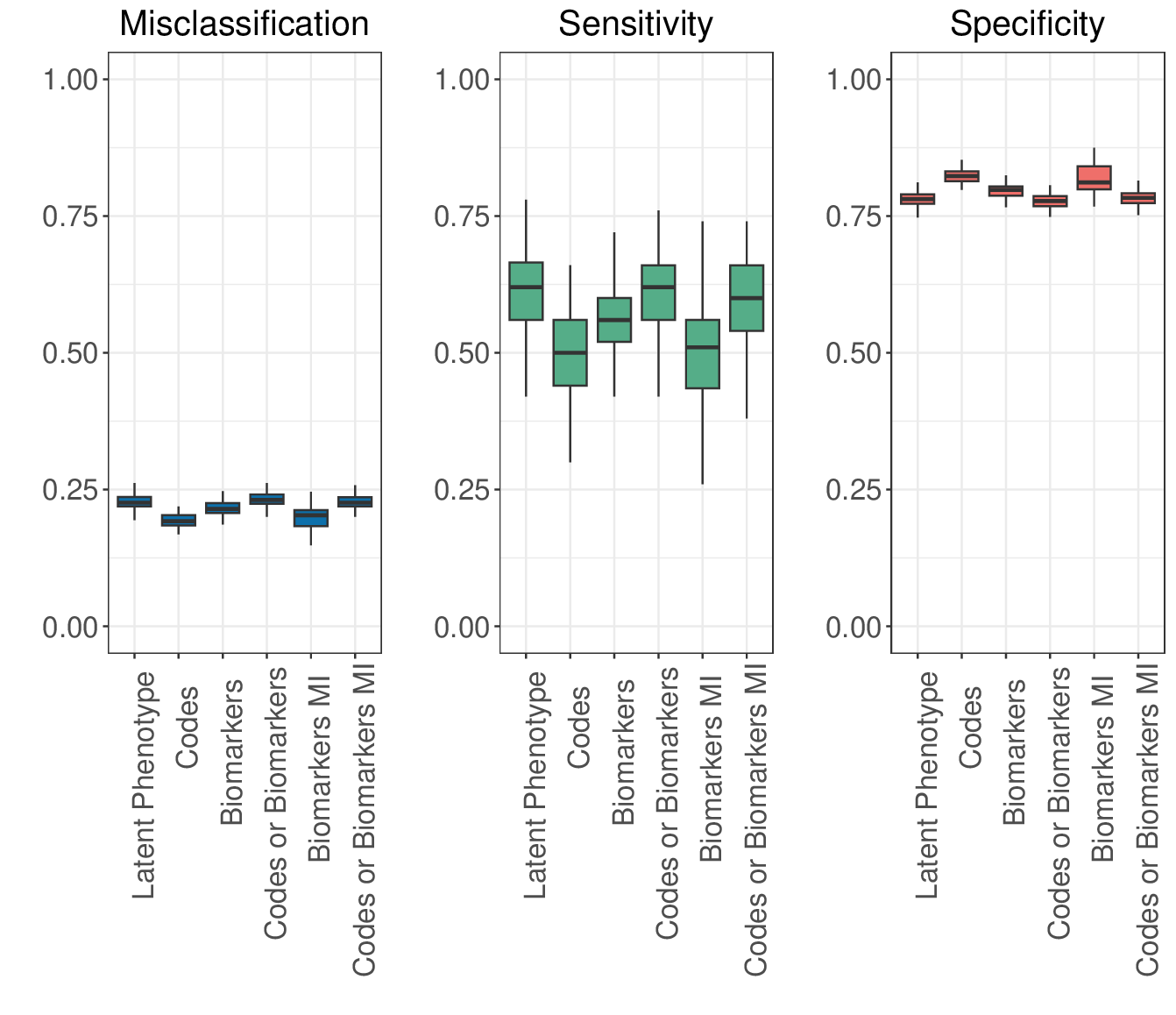}}
  \fbox{\includegraphics[width=0.45\linewidth]{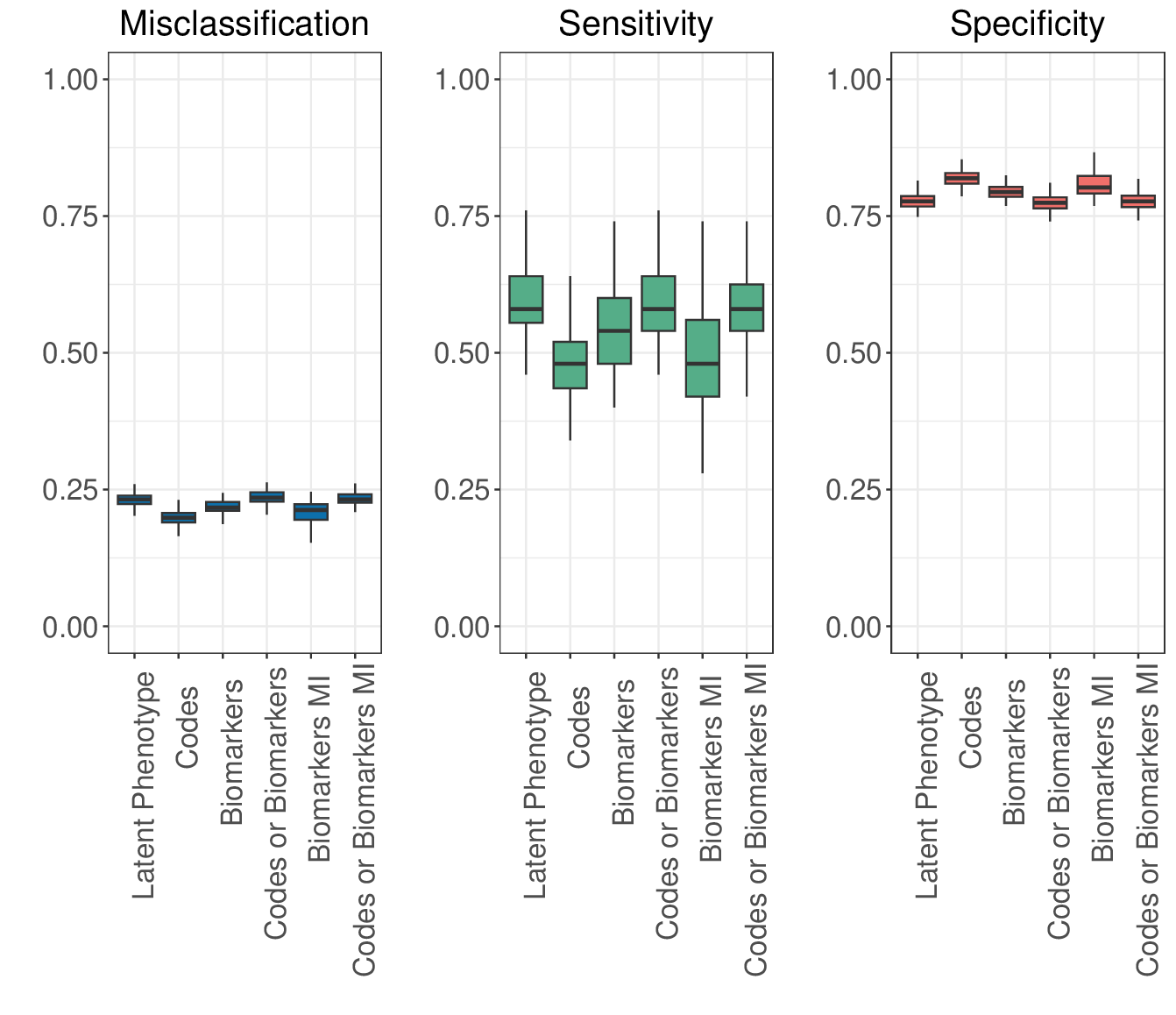}} \\
  \fbox{\includegraphics[width=0.45\linewidth]{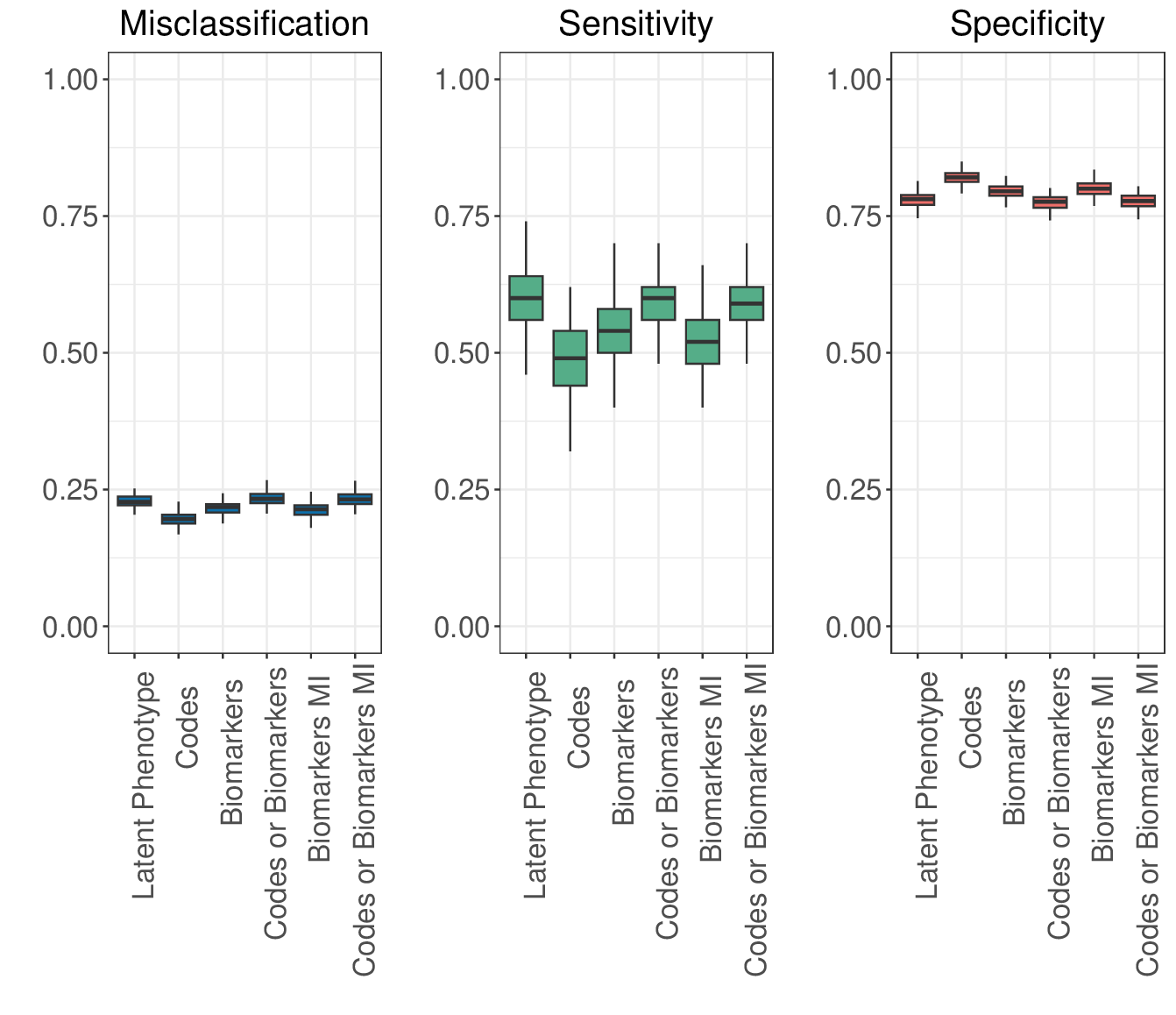}}
  \fbox{\includegraphics[width=0.45\linewidth]{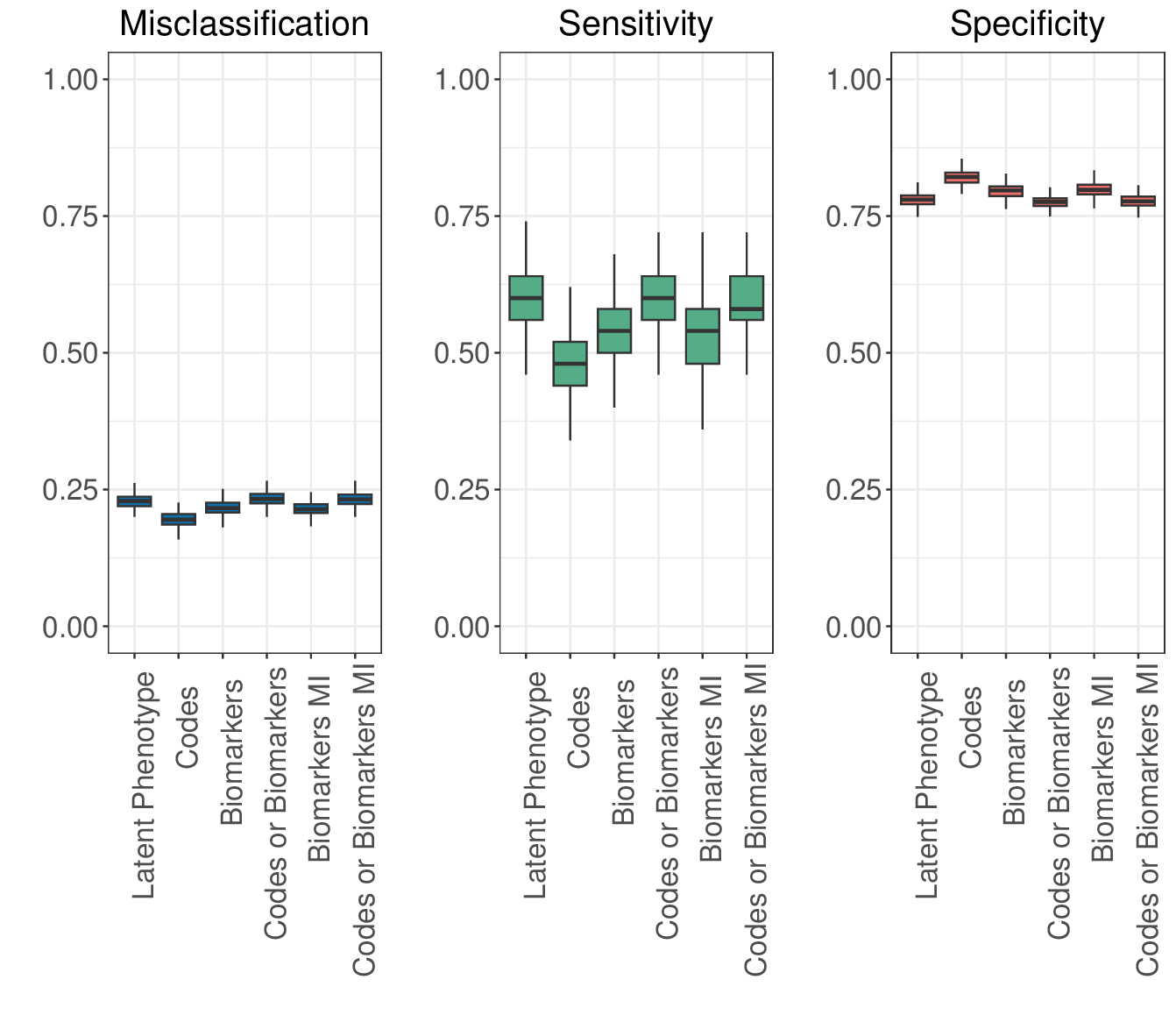}}
\caption{Simulation sensitivity study results for VB comparing high MAR/MNAR (top) and low MAR/MNAR (bottom).}
\label{fig:mle}
\end{figure}

%%% If you don't add the figures in the LaTeX files, please upload them when submitting the article.
%%% Frontiers will add the figures at the end of the provisional pdf automatically
%%% The use of LaTeX coding to draw Diagrams/Figures/Structures should be avoided. They should be external callouts including graphics.

\clearpage
\section*{Tables}
\begin{center}
\captionof{table}{VB and MCMC algorithms in the logistic regression case study.  Automatic VB methods do not require analytical derivation of the ELBO objective function.}
\vspace{3mm}
\scalebox{0.80}{
\begin{tabular}{l l l l l}
\toprule
\textbf{Algorithm} & \textbf{Description} & \textbf{Type} & \textbf{Automatic} & \textbf{Programming} \\
\midrule
CAVI & Coordinate Ascent Variational Inference  & mean-field VB & No & R \\
Own VI & Textbook implementation of CAVI based on~\citep{murphy2012machine} & mean-field VB & No & R \\
SVI & Stochastic Variational Inference  & mean-field VB & No & R \\
varbvs & Fast Variable Selection for Large-scale Regression & mean-field VB & No & R \\
sparsevb & Spike-and-Slab VB for Linear and Logistic Regression & mean-field VB & No & R \\
Stan MC & Stan Hamiltonian Monte Carlo & MCMC & No & R (rstan) \\
Stan VB & Stan ADVI Variational Bayes & mean-field or full-rank VB & Yes & R (rstan, CmdStanR) \\
ADVI & Automatic Differentiation Variational Inference & mean-field VB & Yes & Python \\
FRADVI & Full-rank Automatic Differentiation Variational Inference  & full-rank VB & Yes & Python \\
NFVI &  Normalizing Flow Variational Inference & mean-field or full-rank VB & Yes & Python \\
ASVGD & Amortized Stein Variational Gradient Descent & operator VB & Yes & Python \\
\bottomrule
\label{tab:vb_algos2}
\end{tabular}}
\end{center}

\clearpage
\begin{center}
\captionof{table}{Model specification for Bayesian latent variable model for EHR-derived phenotypes for patient $i$.}
\vspace{5mm}
\scalebox{0.80}{
\begin{tabular}{lllll}
\toprule
                                        & \multicolumn{1}{c}{\bfseries Variable} & \multicolumn{1}{c}{\bfseries Model} & \multicolumn{1}{c}{\bfseries Priors}   \\
\midrule
                                        &                       &                                                                &            \\
\textbf{Latent Phenotype}               & $D_i$                 & $D_i \sim $Bern$(g(\bm{X}_i\bm{\beta}^D + \eta_i))$            & $\beta^D \sim $MVN$(0, \Sigma_D); \eta_i \sim $Unif$(a,b)$              \\
                                        &                       &                                                                &            \\
\textbf{Availability of Biomarkers}     & $R_{ij}, j=1,...,J$   & $R_{ij} \sim $Bern$(g((1,\bm{X}_i,D_i)\bm{\beta}^R_j))$        & $\beta^R_j \sim $MVN$(\mu_R, \Sigma_R)$                                  \\
                                        &                       &                                                                &            \\
\textbf{Biomarkers}                     & $Y_{ij}, j=1,...,J$   & $Y_{ij} \sim $N$(g((1,\bm{X}_i,D_i)\bm{\beta}^Y_j, \tau^2_j))$ & $\beta^Y_j \sim $MVN$(\mu_Y, \Sigma_Y); \tau^2_j \sim $InvGamma$(c,d)$  \\
                                        &                       &                                                                &            \\
\textbf{Clinical Codes}                 & $W_{ik}, k=1,...,K$   & $W_{ik} \sim $Bern$(g((1,\bm{X}_i,D_i)\bm{\beta}^W_k))$        & $\beta^W_k \sim $MVN$(\mu_W, \Sigma_W)$                                  \\
                                        &                       &                                                                &            \\
\textbf{Prescription Medications}       & $P_{il}, l=1,...,L$   & $P_{il} \sim $Bern$(g((1,\bm{X}_i,D_i)\bm{\beta}^P_l))$        & $\beta^P_l \sim $MVN$(\mu_P, \Sigma_P)$                                  \\
                                        &                       &                                                                &            \\
\bottomrule
                                        &                       &                                                                &            \\
                                        &                       &       & \multicolumn{1}{c}{$\bm{ g(\bm{\cdot}) = exp(\bm{\cdot})/(1 + exp(\bm{\cdot})) }$} \\
\end{tabular}}
\label{table:modelSpec2}
\end{center}

\clearpage
\begin{center}
\captionof{table}{Mapping the composite Bayes LCA/regression phenotyping model factors to the pediatric T2DM study for patient $i$. }
\vspace{3mm}
\scalebox{0.80}{
\begin{tabular}{lllll}
\toprule
                                        & \multicolumn{1}{c}{\bfseries Model Variable} & \multicolumn{1}{c}{\bfseries Data Elements}                \\
\midrule
                                        &                       &                                                                                   \\
\textbf{Latent Phenotype}               & $D_i$                 & Presence of T2DM for observation $i$ (binary latent variable, $D_i \in \{0,1\}$,  \\
                                        &                       & where $1$ indicates presence of T2DM) based on patient characteristics            \\
                                        &                       & e.g. demographics                                                                 \\
                                        &                       &                                                                                   \\
\textbf{Availability of Biomarkers}     & $R_{ij}, j=1,...,J$   & Availability of ($j$=1) Glucose and ($j$=2) HbA1c biomarker data                  \\
                                        &                       &                                                                                   \\
\textbf{Biomarkers}                     & $Y_{ij}, j=1,...,J$   & Laboratory test values for ($j$=1) Glucose and ($j$=2) HbA1c  CPT codes           \\
                                        &                       &                                                                                   \\
\textbf{Clinical Codes}                 & $W_{ik}, k=1,...,K$   & ($k$=1) ICD Code for T2 Diabetes and ($k$=2) CPT code for Endocrinologist Visit   \\
                                        &                       &                                                                                   \\
\textbf{Prescription Medications}       & $P_{il}, l=1,...,L$   & NCD Medication codes for ($l$=1) Insulin and ($l$=2) Metformin                    \\

                                        &                       &                                                                                   \\
\bottomrule
\label{table:modelExample2}
\end{tabular}}
\end{center}

\clearpage
\begin{center}
\captionof{table}{Comparison of Optum\textsuperscript{\texttrademark} results with Hubbard et al. PEDSnet using the same JAGS LCA model published in Hubbard et al. GitHub~\citep{hubbard2019bayesian})}
\vspace{3mm}
\scalebox{0.80}{
\begin{tabular}{llll}
\toprule
\textbf{}                               & \multicolumn{2}{l}{\textbf{Posterior Mean (95\% CI)}}                   \\
                                        & (a) PEDSnet data                   & (b) Optum\textsuperscript{\texttrademark} data \\
\midrule
                                        &                                    &                                    \\
T2DM code sensitivity (expit($\beta^W_{10} + \beta^W_{11}$))                   & 0.17 (0.15, 0.20)       &  0.15 (0.12, 0.18)      \\[1mm]
T2DM code specificity (1-expit($\beta^W_{10}$))                   & 1.00 (1.00, 1.00)       &  1.00 (1.00, 1.00)                   \\[1mm]
Endocrinologist visit code sensitivity (expit($\beta^W_{20} + \beta^W_{21}$))  & 0.94 (0.92, 0.95)       &  0.18 (0.15, 0.21)      \\[1mm]
Endocrinologist visit code specificity (1-expit($\beta^W_{20}$))  & 0.93 (0.93, 0.94)       &  0.99 (0.98, 0.99)                   \\[1mm]
Metformin code sensitivity (expit($\beta^P_{10} + \beta^P_{11}$))              & 0.31 (0.28, 0.35)       &  0.40 (0.36, 0.44)      \\[1mm]
Metformin code specificity (1-expit($\beta^P_{10}$))              & 0.99 (0.99, 0.99)       &  0.98 (0.98, 0.99)                   \\[1mm]
Insulin code sensitivity (expit($\beta^P_{20} + \beta^P_{21}$))                & 0.66 (0.61, 0.70)       &  0.55 (0.51, 0.59)      \\[1mm]
Insulin code specificity (1-expit($\beta^P_{20}$))                & 1.00 (1.00, 1.00)       &  1.00 (1.00, 1.00)                   \\[1mm]
Mean shift in HbA1c ($\beta^Y_{12}$)                     & 3.15 (3.06, 3.24)       &  4.80 (4.72, 4.81)                            \\[1mm]
Mean shift in Glucose ($\beta^Y_{11}$)                   &            90.62 (90.25, 91.00)    &            89.30 (89.10, 90.01)    \\[1mm]
\bottomrule
\label{table:comp2}
\end{tabular}}
\end{center}

\clearpage
\begin{center}
\captionof{table}{Simulation results for VB low MAR}
\vspace{3mm}
\scalebox{0.80}{
\begin{tabular}{lllll}
\toprule
\textbf{}           & \multicolumn{2}{l}{\textbf{Posterior Median (95\% CI)}} \\
& Misclassification  & Sensitivity & Specificity \\
\midrule
& & & \\
\multicolumn{3}{l}{\textbf{VB Low MAR}}  &  \\
& & & \\
Latent Phenotype & 0.23 (0.20,0.26) & 0.60 (0.49,0.72) & 0.78 (0.75,0.80) \\[1mm]
Codes & 0.20 (0.17,0.22) & 0.49 (0.36,0.60) & 0.82 (0.79,0.85) \\[1mm]
Biomarkers & 0.22 (0.20,0.25) & 0.54 (0.42,0.67) & 0.80 (0.77,0.82) \\[1mm]
Codes or Biomarkers & 0.23 (0.21,0.27) & 0.60 (0.48,0.70) & 0.78 (0.74,0.80) \\[1mm]
Biomarkers MI & 0.21 (0.19,0.25) & 0.52 (0.40,0.65) & 0.80 (0.77,0.82) \\[1mm]
Codes or Biomarkers MI & 0.23 (0.21,0.27) & 0.59 (0.48,0.70) & 0.78 (0.75,0.80) \\[1mm]
& & & \\
\multicolumn{3}{l}{\textbf{VB Low MNAR}}  &  \\
& & & \\
Latent Phenotype & 0.23 (0.20,0.25) & 0.60 (0.48,0.72) & 0.78 (0.75,0.81) \\[1mm]
Codes & 0.19 (0.16,0.22) & 0.48 (0.37,0.60) & 0.82 (0.80,0.85) \\[1mm]
Biomarkers & 0.22 (0.19,0.24) & 0.54 (0.40,0.67) & 0.80 (0.77,0.83) \\[1mm]
Codes or Biomarkers & 0.23 (0.20,0.26) & 0.60 (0.46,0.72) & 0.78 (0.75,0.81) \\[1mm]
Biomarkers MI & 0.21 (0.18,0.24) & 0.54 (0.39,0.63) & 0.80 (0.78,0.83) \\[1mm]
Codes or Biomarkers MI & 0.23 (0.20,0.26) & 0.58 (0.45,0.72) & 0.78 (0.75,0.81) \\[1mm]
& & & \\
\multicolumn{3}{l}{\textbf{VB High MAR}}  &  \\
& & & \\
Latent Phenotype & 0.23 (0.20,0.25) & 0.62 (0.48,0.75) & 0.78 (0.76,0.81) \\[1mm]
Codes & 0.19 (0.17,0.22) & 0.50 (0.36,0.64) & 0.82 (0.80,0.84) \\[1mm]
Biomarkers & 0.21 (0.19,0.24) & 0.56 (0.38,0.69) & 0.80 (0.77,0.82) \\[1mm]
Codes or Biomarkers & 0.23 (0.20,0.26) & 0.62 (0.46,0.73) & 0.78 (0.75,0.80) \\[1mm]
Biomarkers MI & 0.20 (0.16,0.24) & 0.51 (0.30,0.65) & 0.81 (0.78,0.87) \\[1mm]
Codes or Biomarkers MI & 0.23 (0.20,0.26) & 0.60 (0.45,0.72) & 0.78 (0.75,0.81) \\[1mm]
& & & \\
\multicolumn{3}{l}{\textbf{VB High MNAR}}  &  \\
& & & \\
Latent Phenotype & 0.23 (0.20,0.26) & 0.58 (0.47,0.74) & 0.78 (0.75,0.81) \\[1mm]
Codes & 0.20 (0.17,0.22) & 0.48 (0.34,0.63) & 0.82 (0.80,0.85) \\[1mm]
Biomarkers & 0.22 (0.19,0.24) & 0.54 (0.42,0.70) & 0.79 (0.77,0.82) \\[1mm]
Codes or Biomarkers & 0.23 (0.20,0.26) & 0.58 (0.47,0.73) & 0.77 (0.75,0.80) \\[1mm]
Biomarkers MI & 0.21 (0.16,0.24) & 0.48 (0.34,0.68) & 0.80 (0.77,0.86) \\[1mm]
Codes or Biomarkers MI & 0.23 (0.20,0.26) & 0.58 (0.46,0.72) & 0.78 (0.75,0.81) \\[1mm]
& & & \\
\bottomrule
\label{table:sim1}
\end{tabular}}
\end{center}

\clearpage
\begin{center}
\captionof{table}{Comparison of composite LCA/regression model results for clinical attributes}
\vspace{3mm}
\scalebox{0.80}{
\begin{tabular}{lllll}
\toprule
\textbf{}                               & \multicolumn{3}{l}{\textbf{Posterior Mean (95\% CI)}}                                                     \\
                                        & (a) JAGS Gibbs MCMC           & (b) Stan HMC           & (c) Stan VB                                      \\
\midrule
                                        &                         &                       &                                                         \\
T2DM code sensitivity (expit($\beta^W_{10} + \beta^W_{11}$))                   &  0.15 (0.12, 0.18)      &  0.10 (0.09, 0.11)    & 0.12 (0.10, 0.12)      \\[1mm]
T2DM code specificity (1-expit($\beta^W_{10}$))                     &  1.00 (1.00, 1.00)      &  1.00 (0.99, 1.00)    & 0.99 (0.99, 0.99)      \\[1mm]
Endocrinologist visit code sensitivity (expit($\beta^W_{20} + \beta^W_{21}$))   &  0.18 (0.15, 0.21)      &  0.20 (0.18, 0.21)    & 0.22 (0.19, 0.22)      \\[1mm]
Endocrinologist visit code specificity (1-expit($\beta^W_{20}$))  &  0.99 (0.98, 0.99)      &  0.98 (0.97, 0.99)    & 0.97 (0.97, 0.99)      \\[1mm]
Metformin code sensitivity (expit($\beta^P_{10} + \beta^P_{11}$))              &  0.40 (0.36, 0.44)      &  0.21 (0.20, 0.21)    & 0.19 (0.19, 0.20)      \\[1mm]
Metformin code specificity (1-expit($\beta^P_{10}$))               &  0.98 (0.98, 0.99)      &  0.93 (0.92, 0.93)    & 0.93 (0.92, 0.94)      \\[1mm]
Insulin code sensitivity (expit($\beta^P_{20} + \beta^P_{21}$))                &  0.55 (0.51, 0.59)      &  0.35 (0.31, 0.35)    & 0.20 (0.19, 0.20)      \\[1mm]
Insulin code specificity (1-expit($\beta^P_{20}$))                &  1.00 (1.00, 1.00)      &  1.00 (0.99, 1.00)    & 1.00 (0.99, 1.00)      \\[1mm]
Mean shift in HbA1c ($\beta^Y_{12}$)                     &  4.80 (4.72, 4.81)      &  4.77 (4.76, 4.78)    & 4.77 (4.75, 4.78)      \\[1mm]
Mean shift in glucose ($\beta^Y_{11}$)                   & 89.30 (89.10, 90.01)                                     & 88.59 (88.48, 88.71)                                   & 22.80 (21.06, 24.92) \\
\bottomrule
\label{table:allResults2}
\end{tabular}}
\end{center}

\clearpage
\begin{table}[ht]
\centering
\captionof{table}{Resulting coefficients for 5 folds for BloodPressure and Insulin comparing MCMC with Stan\_MC and Stan\_VB}
\vspace{3mm}
\begin{tabular}{lrrr}
\toprule
\textbf{} Method & fold & BloodPressure & Insulin \\
\midrule
 MCMC & 1 & -0.017 & -0.002 \\
 MCMC & 2 & -0.011 & -0.001 \\
 MCMC & 3 & -0.011 & -0.001 \\
 MCMC & 4 & -0.011 & -0.001 \\
 MCMC & 5 & -0.015 & -0.001 \\
 Stan\_MC & 1 & 0.020 & 0.002 \\
 Stan\_MC & 2 & 0.020 & 0.002 \\
 Stan\_MC & 3 & 0.020 & 0.002 \\
 Stan\_MC & 4 & 0.020 & 0.002 \\
 Stan\_MC & 5 & 0.020 & 0.002 \\
 Stan\_VB & 1 & 0.020 & 0.006 \\
 Stan\_VB & 2 & 0.009 & 0.010 \\
 Stan\_VB & 3 & 0.030 & 0.008 \\
 Stan\_VB & 4 & 0.030 & -0.009 \\
 Stan\_VB & 5 & 0.030 & 0.006 \\
\bottomrule
\label{table:allResults3}
\end{tabular}
\end{table}

\begin{table}[ht]
\centering
\captionof{table}{Proportions of zero values in Pima Indians data in total, for the negative Outcome class and for the positive Outcome class}
\vspace{3mm}
\begin{tabular}{lrrr}
\toprule
\textbf{} Variable & Total & Negative & Positive \\
\midrule
Pregnancies & 0.145 & 0.146 & 0.142 \\
Glucose & 0.007 & 0.006 & 0.007 \\
BloodPressure & 0.046 & 0.038 & 0.060 \\
SkinThickness & 0.296 & 0.278 & 0.328 \\
Insulin & 0.487 & 0.472 & 0.515 \\
BMI & 0.014 & 0.018 & 0.007 \\
\bottomrule
\label{table:allResults4}
\end{tabular}
\end{table}

\end{document}